\documentclass[a4paper,12pt]{article}

\usepackage[dvips]{graphicx}
\usepackage{cite}
\usepackage{pstricks}
\usepackage{multirow}
\usepackage{latexsym}
\usepackage{amssymb}
\usepackage{bm}

\usepackage{graphicx}

\newcommand{\be}{\begin{equation}}
\newcommand{\ee}{\end{equation}}

\newcommand{\beq}[1] {\begin{equation}\label{#1} }
\newcommand{\eeq} {\end{equation} }
\newcommand{\bea}[1]{\begin{eqnarray}\label{#1} }
\newcommand{\eea}{\end{eqnarray}}

\def\beqn{\begin{eqnarray}}
\def\eeqn{\end{eqnarray}}

\def\beq{\begin{equation}}
\def\eeq{\end{equation}}
\def\bea{\begin{equation}}
\def\eea{\end{equation}}

\def\hs{\hspace}
\def\vs{\vspace}

\begin{document}
\vspace*{-0.2in}
\begin{flushright}
OSU-HEP-09-04\\
July 24, 2009\\
\end{flushright}

\vs{0.5cm}

\begin{center}
{\Large\bf Flavor Violation in Supersymmetric $Q_6$ Model}\\
\end{center}

\vspace{0.5cm}
\begin{center}
{\large
{}~K.S. Babu\footnote{E-mail: babu@okstate.edu}{} and
{}~Yanzhi Meng\footnote{E-mail: yanzhi.meng@okstate.edu}
}
\vspace{0.5cm}

{\em Department of Physics \\
Oklahoma State University\\
Stillwater, OK 74078, USA }
\end{center}

\begin{abstract}

We investigate flavor violation mediated by Higgs bosons and supersymmetric particles in a
predictive class of models based on the non--Abelian flavor symmetry $Q_6$.  These models, which aim to
reduce the number of parameters of the fermion sector and to solve the flavor changing problems of generic SUSY setup, assume three families of Higgs bosons and spontaneous/soft violation of $CP$ symmetry.
Tree--level contributions to meson--antimeson mixings mediated by Higgs bosons are shown to be within experimental
limits for Higgs masses in the (1-5) TeV range.  Calculable flavor violation induced by SUSY loops are
analyzed for meson mixing and lepton decays and found to be consistent with data.  Significant new
SUSY contributions arise in $B_s- \overline{B}_s$ mixing, but non-standard $CP$ violation is suppressed. A simple solution to the SUSY $CP$ problem is found, which requires light Higgsinos.

\end{abstract}

\newpage

\section{Introduction}

     The gauge interactions of the standard model (SM) fermions are invariant under separate $U(3)_L \times U(3)_R$ transformations.  This global symmetry is broken explicitly by the fermion Yukawa couplings.  For the
light fermions violation of this symmetry is small, being proportional to their masses.  This feature has
played a crucial role in the success of the SM in the flavor sector.  In extensions of the SM this property is generally lost, often leading to excessive flavor changing neutral current (FCNC) processes.

A case in point is the supersymmetric standard model which is the subject of this paper. While the gauge interactions of the SUSY SM respect
the $U(3)_L \times U(3)_R$ global symmetry, there are new sources of violation of this symmetry, in
the soft SUSY breaking sector.  Indeed, generic soft SUSY breaking scenarios lead to excessive FCNC in processes
such ad $K^0-\overline{K^0}$ mixing, $B^0-\overline{B^0}$ mixing, $D^0-\overline{D^0}$ mixing, and flavor changing leptonic decays such as $\mu \rightarrow e\gamma$ \cite{susyfcnc}.
This problem is most severe in the $K^0-\overline{K^0}$ system.   SUSY box diagrams involving
gluino and squarks modify the successful SM prediction for $\Delta M_K$ and $\epsilon_K$, leading to the
following constraints for the real and imaginary parts of the amplitude \cite{ciuchini}:
\begin{equation}\label{limit}
\left|({\rm Re,~Im})(\delta^d_{LL})_{12} (\delta^d_{RR})_{12}\right|^{1/2} \leq (9.6 \cdot 10^{-4},~1.3 \cdot 10^{-4}) \left({\tilde{m} \over 500 ~{\rm GeV}}\right)~.
\end{equation}
Here $(\delta_{AB})_{ij} = (m^2_{AB})_{ij}/\tilde{m}^2$ is a flavor violating squark mass insertion parameter,
for $(A,B) = (L, R)$, with $\tilde{m}$ being the average mass of the relevant squarks $(\tilde{d}$
and $\tilde{s}$ in this case).  For this estimate the gluino mass was assumed to equal the average squark mass.
Now, the natural magnitude of the mixing parameters $(\delta^d_{LL})_{12}$ and $(\delta^d_{RR})_{12}$, in the
absence of additional symmetries, should be of order the Cabibbo angle, $\sim 0.2$.  Since the parameters
$(\delta_{AB})_{ij}$ split the masses of the squarks, one sees from Eq. (\ref{limit}) that a high degree of
squark mass degeneracy is needed for consistency.

Analogous limits from $B_d^0-\overline{B_d^0}$ mixing are less severe, as given by \cite{becirevic}:
\begin{equation}
\left|({\rm Re,~Im})(\delta^d_{LL})_{13} (\delta^d_{RR})_{13}\right|^{1/2} \leq (2.1 \cdot 10^{-2},~9.0 \cdot 10^{-3}) \left({\tilde{m} \over 500 ~{\rm GeV}}\right)~. \label{Bfcnc}
\end{equation}
Note that the natural value of this mixing parameter, in the absence of other symmetries, is $V_{ub} \sim
3 \times 10^{-3}$.  The constraints from Eq. (\ref{Bfcnc}) are well within limits.  $B_s-\overline{B}_s$ mixing
provides even weaker constraints.

It can be argued that a natural explanation for solving this problem is to enhance the symmetry of the
SUSY SM by assuming a non--Abelian symmetry $G$ (a subgroup of the $U(3)_L \times U(3)_R$) that pairs the first two families into a doublet, with the
third family transforming trivially  \cite{dine}.\footnote{Grouping all three families into an irreducible triplet representation of $G$ is also possible. The large top quark mass however reduces the original $U(3)_L \times U(3)_R$ symmetry to $U(2)_L \times U(2)_R$, so we find it is easier to work with $(2+1)$ assignment.}   Invariance under $G$ will then lead to
degeneracy of squarks, as needed for phenomenology.   A variety of such models have been
proposed in the literature \cite{dine}, \cite{local}, \cite{seiberg}, \cite{babu}, \cite{kubolenz}.
In Ref. \cite{dine}, $SU(2)$ family symmetry and its variants were proposed to solve the SUSY FCNC problem.  If the symmetry
is global, one has to deal with the Goldstone bosons associated with its spontaneous breaking.  Global
symmetries are susceptible to violations from quantum gravity.
Local gauge symmetries are more natural, but in the SUSY context there would be new FCNC processes
arising from the family $SU(2)$ $D$--terms \cite{kawamura}.  Exceptions to this generic problem are known to exist \cite{local}.

A more natural solution to the problem is perhaps to choose $G$ to be a non--Abelian discrete symmetry group \cite{seiberg}.
In this case there would be no $D$--term problem, since there are no gauge bosons associated with $G$.  Spontaneous breaking of such symmetries will not lead to
Goldstone bosons.  If the symmetry breaking occurs before the inflationary era, such models should also be
safe from potential cosmological domain wall problems. Such non--Abelian discrete symmetries have found application in understanding the various puzzles associated with the quark and lepton masses and mixing angles with or without
supersymmetry\cite{other}, more
recently for understanding the tri--bimaximal neutrino mixing pattern \cite{othernu}.  It would be desirable
to find a symmetry that sheds light on the fermion mass and mixing puzzle, and at the same time solves
the SUSY FCNC problem.

The supersymmetric standard model has another problem. In the flavor conserving sector CP violation is
generically too large.  Neutron and electron electric dipole moments (EDM) receive new contributions
from SUSY loops.  Unless the new phases in the SUSY breaking sector are small or conspire to be small,
experimental limits on the EDM of the neutron $(d_n)$, electron $(d_e)$, and
atoms will be violated by two to three orders of magnitude (depending on the squark and
slepton masses) \cite{edm}, \cite{nath}.
The imaginary parts of the left--right squark mixing parameters must satisfy the constraints
(from the experimental constraints $d_n < 6.3 \times 10^{-26}$ e-cm, $d_e < 4.3 \times 10^{-27}$ e-cm) \cite{abel}
\begin{equation}
{\rm Im}[(\delta^d_{LR})_{11}] \leq 1.9 \times 10^{-6} \left({\tilde{m} \over 500 ~{\rm GeV}}\right),~~~
{\rm Im}[(\delta^e_{LR})_{11}] \leq 1.7 \times 10^{-7} \left( {\tilde{m} \over 100 ~{\rm GeV}}  \right),
\end{equation}
assuming that the gluino/Bino has the same mass as the squark/slepton.  Now, since these mixing parameters
are expected to be suppressed by fermion helicity factors (but enhanced by the MSSM parameter $\tan\beta$)
the natural values for these mixing parameters are of order $(1 \times 10^{-4},~3 \times 10^{-6})$ respectively,
(for $\tan\beta = 10$ and assuming order one phases).  This implies that the CP violating phases arising from
the soft SUSY breaking sector must satisfy $\theta_d \leq 1/53$, $\theta_e \leq 1/63$ (for gluino (Bino) mass
of 500 GeV (100 GeV).  Why this is so, while the Kobayashi-Maskawa phase takes order one value, is the SUSY
CP puzzle. It would be desirable to resolve this puzzle based on a symmetry principle in the same context where
the SUSY FCNC problem is solved.

The purpose of this paper is to study  a recently proposed
SUSY model based on the non--Abelian symmetry group $Q_6$ \cite{babu} which
addresses these issues.  $Q_6$ is a finite subgroup of $SU(2)$
with twelve elements.  Apart from providing a solution to the SUSY flavor problem, this class
of models can also constrain the quark masses and mixings.  It was shown in Ref. \cite{babu} that with
the assumption of spontaneous (or soft) CP violation, there is a non-trivial relation between quark
masses and mixings in this model.  This sum rule was found to be consistent with experimental data.

A crucial aspect of the $Q_6$ model relevant for the quark mixing sum rule is that CP violation occurs either spontaneously or softly.  This can help ameliorate the SUSY CP problem mentioned above.  CP invariance requires that the gaugino masses, the $\mu$ terms and the trilinear $A$ terms be all real. In the $Q_6$ model of Ref. \cite{babu}
it was found that there is a phase alignment mechanism that makes the phases of the sfermion mixing terms arising from the $A$--terms to align with the phases of the
fermion masses.  So SUSY CP violation is suppressed to a large extent.   However, spontaneously induced complex VEVs
do lead to non-zero contributions to EDM.  Here we analyze these contributions.  Since these complex VEVs
are accompanied by the Higgsino $\mu$ terms, a simple solution to the problem is found by making the Higgsinos
to be lighter than the squarks.  Adequate suppression of EDM is obtained for $\mu \sim 100$ GeV, while
squark masses are of order 500 GeV.  This suggestion obviously has testable implications for physics that will
be probed at the LHC.

The fermion mass matrices that allow for a non-trivial prediction and the phase alignment is a generalization
of well studied models \cite{weinberg}.  The mass matrices for up and down quarks and the charged leptons
take the following form:
\begin{equation}\label{massmatrix}
\mathbf{M}=\left(\begin{array}{ccc}0&C&0\\ \pm C&0&B\\0&B'&A \end{array}\right)\label{m0}~.
\end{equation}
The main feature of such mass matrices is that the phases can be factorized, i.e., $M = P \cdot M^0 \cdot Q$, with
$M^0$ being real and $P,Q$ being diagonal phase matrices.  This feature, when combined with the $Q_6$ symmetry,
has an the interesting consequence
that CP violation induced by SUSY loops are suppressed.  This will be discussed in more detail in Sec. 4.

The form of Eq. (\ref{m0}) can be obtained in renormalizable theories based on  $Q_6$ symmetry.
This requires the introduction of three families of Higgs doublets, which fall into $2+1$ representations
of the $Q_6$ group, very much like the quarks and leptons.  With multiple Higgs fields coupling to
fermions, invariably there will be tree-level FCNC mediated by the Higgs bosons.  The flavor changing Higgs
couplings are not arbitrary, but can be computed in terms of the fermion masses and mixings.  We will show that these
FCNC processes are within acceptable range, provided that the Higgs boson masses lie in the $(1-5)$ TeV range
(except of course for the standard model--like Higgs boson, which has a mass in the $(100 - 130)$ GeV range).
While Higgsinos are naturally light in this scenario, in the bosonic sector only the lightest SM--like
Higgs will be accessible to LHC experiments.

One of our major results is that non--standard CP violation is highly suppressed in this class of models.
The phase factorizability of the fermion mass matrices implies that much of the SUSY induced CP violation
is small.  The structure of the Yukawa couplings in the model implies that the amplitudes for
tree--level FCNC induced by neutral Higgs bosons are nearly real (see discussions in Sec. 5).  While there can be
significant new contributions to meson--antimeson mixings, there is very little CP violation beyond
the standard model.

Our analysis is similar in spirit to that of Ref. \cite{kubolenz}.  Our approach is slightly different,
with some differences in analytical results, fits, spectrum, and conclusions. In particular, we have presented complete analytical results for the Higgs boson spectrum, and we have a new proposal to solve the SUSY EDM problem, which requires light Higgsinos.  We have also derived generalized constraints on SUSY FCNC parameters for the
$B_{d,s}-\overline{B}_{d,s}$ system appropriate for a (2+1) mass spectrum.

The plan of the paper is as follows.  In Sec. 2 we describe the SUSY $Q_6$ model, lay out the parameter
choice, and summarize the prediction for the quark sector.  In Sec. 3 we analyze the Higgs potential
involving the three pairs of Higgs doublets.  We provide analytic expressions for the mass spectrum of
Higgs bosons as well as numerical fits.  Consistency of symmetry breaking and spontaneous CP violation
will be established here.  In Sec. 4 we address tree--level FCNC processes mediated by the
heavy Higgs bosons.  Sec. 5 is devoted to analysis of the SUSY flavor violation and EDM within the model.
In Sec. 6 we conclude.

\section{Supersymmetric $Q_6$ Model}

$Q_6$ is the binary dihedral group, a subgroup of $SU(2)$, of order 12.  It has the presentation
\begin{equation}
\{A, B; A^6=E, B^2=A^3, B^{-1}AB=A^{-1}\}~.
\end{equation}
The 12 elements of $Q_6$ can be represented as
\begin{equation}
\{ E, A, A^2, ... , A^5, B, BA, BA^2, ..., BA^5\}.
\end{equation}
In the two dimensional representation the generators are given in a certain basis by
\begin{equation}
  \mathbf{A}=\left(\begin{array}{cc}\cos{\frac{\pi}{3}}&\sin{\frac{\pi}{3}}\\-\sin{\frac{\pi}{3}}&\cos{\frac{\pi}{3}}
  \end{array}\right)
  \hspace{1cm}  \mathbf{B}=\left( \begin{array}{cc} i&0\\0&-i \end{array}\right)~.
\end{equation}
The irreducible representation of
$Q_6$ fall into $2, 2^{'}, 1, 1^{'}, 1^{''}, 1^{'''}$,
where the 2 is complex--valued but pseudoreal, while the $2^{'}$ is real valued. ($Q_6$ is the simplest
group with two distinct doublet representations, which is very useful for model building.)
The 1 and $1^{'}$ are real representations, while $1^{''}$ and $1^{'''}$ are
complex conjugates to each other. The group multiplication rules are given as
\begin{equation} 1'\times 1'=1,\ 1''\times 1''=1,\ 1'''\times 1'''=1',\ 1''\times 1'''=1,\
1'\times1'''=1'',\ 1\times1''=1''' \end{equation}
\begin{equation} 2\times1'=2,\ 2\times1''=2',\ 2\times1'''=2',\ 2'\times1'=2',\ 2'\times1''=2,\ 2'\times1'''=2
\end{equation}
\begin{equation} 2\times2=1+1'+2',\ 2'\times2'=1+1'+2',\ 2\times2'=1''+1'''+2 \end{equation}
The Clebsch--Gordon coefficients for these multiplication can be found in Ref. \cite{babu}.

The fermions of all sectors (up--quark, down--quark, charged leptons) are assigned to $2+1$ represtations
of $Q_6$.  The model assumes three families of Higgs bosons, which are also assigned to $2+1$ under $Q_6$.
Their transformation properties are given by
\begin{equation}\psi=\left(\psi_1\atop \psi_2 \right)=2, \quad \psi^c=\left(-\psi^c_1\atop \psi^c_2 \right)=2',
\quad \psi_3=1' \quad \psi^c_3=1''', \end{equation}
\begin{equation} \quad H=\left(H_1\atop H_2  \right)=2', \quad H_3=1'''. \end{equation}
Here $\psi$ generically denotes the fermion fields, and $H$ denotes the up--type and the down--type
Higgs fields which are doublets of $SU(2)_L$.  Due to the constraints of supersymmetry, $H^u$ and
$H_3^u$ couple only to up quarks, while $H^d$ and $H_3^d$ couple to down--type quarks and leptons.
The Yukawa couplings of the model in the down quark sector arise from the superpotential
\begin{equation}
W =\alpha_d\psi_3\psi^c_3H_3+\beta_d\psi^T\tau_1\psi^c_3H-\beta'_d\psi_3{\psi^c}^{T}i\tau_2H
+\delta_d\psi^T\tau_1\psi^c H_3+{\rm h. c.} \label{yuk}
\end{equation}
with similar results for up--type quarks and charged leptons.
This leads to the mass matrix for the down quarks given by
\begin{equation}\mathbf{M_d}=\left( \begin{array}{ccc}0&\delta_d v_{d3}
 &\beta_d v_{d2} \\-\delta_d v_{d3} &0&\beta_d v_{d1} \\\beta'_d v_{d2}
 &\beta'_d v_{d1} &\alpha_d v_{d3}
\end{array}\right)~.\label{Md}
\end{equation}
Here $v_{d1},~v_{d2},~v_{d3}$ are the vacuum expectation values of $H^d_{1,2,3}$ fields, which break
the $Q_6$ symmetry.

Now, the potential of the $Q_6$ model admits an unbroken $S_2$ symmetry which interchanges
$H_1^{u,d} \leftrightarrow H_2^{u,d}$.  This unbroken symmetry allows us to choose a VEV pattern
\begin{equation}
v_{u1}=v_{u2},\hspace{0.5cm}
v_{d1}=v_{d2}, \hspace{0.5cm}~.
\end{equation}
Consequently, a 45$^0$ rotation of the matrix in Eq. (\ref{Md}) in the 1-2 plane can be done
both in the up and the down quark sectors without inducing CKM mixing.  This
will bring the mass matrices to the desired form
of Eq. (\ref{massmatrix}).  By using the unbroken $S_2$ symmetry, we make a 45$^0$ rotation
on the Higgs fields, $\hat{H}_{1,2} = (H_1\pm H_2)/\sqrt{2}$, so that  $\hat{H}_1$
acquires a VEV, while $\langle \hat{H}_2 \rangle = 0$.  We shall drop the hat on these redefined fields, and simply denote the VEV
of the redefined $H_1$ as $v_1$.

We assume that CP is a good symmetry of the Lagrangian, and that it is broken spontaneously by the VEVs
of scalar fields.  If the full theory contains SM singlet Higgs fields, spontaneous CP violation in
the singlet sector will show up as soft CP violation in the Higgs doublet sector.  Explicit examples of
this sort have been given in Ref. \cite{babu}.  For now we simply assume that the Yukawa couplings in
Eq. (\ref{yuk}) are real, and the CKM CP violation has a spontaneous origin, via complex VEVs of the
Higgs doublet fields.  We denote the phase of these (redefined) VEVs as
\begin{equation}
\Delta \theta_u=\arg (v_{u3})-\arg (v_{u1}),
\hspace{1cm}
\Delta \theta_d=\arg (v_{d3})-\arg (v_{d1}).
\end{equation}

We make an overall  $45^{\circ}$ rotation on the $Q_6$ doublets, $Q$, $D^c$ and $U^c$, and then a phase rotations on these fields:
\begin{equation}
U\rightarrow P_u U, \hspace{1cm} U^c\rightarrow P_{u^c}U^c
\end{equation}
and similarly for $D$ and $D^c$ fields, where
\begin{eqnarray}
P_{u,\ d}=\left( \begin{array}{ccc}1&0&0\\0&\exp(i2\Delta\theta_{u,\ d})&0\\0&0&\exp(i\Delta
\theta_{u,\ d})\end{array}\right),\nonumber\\
P_{u^c,\ d^c}=\left( \begin{array}{ccc}\exp(-i2\Delta\theta_{u,\ d})&0&0\\0&1&0\\0&0&\exp(-i\Delta
\theta_{u,\ d})\end{array}\right)~.\label{p1}
\end{eqnarray}
This will make the originally complex mass matrices of Eq. (\ref{massmatrix}) real, which
we parametrize as
\begin{equation}
M_{u,\ d}=m^0_{t,\ b}\left( \begin{array}{ccc} 0&q_{u,\ d}/y_{u,\ d}&0\\
-q_{u,\ d}/y_{u,\ d}&0&b_{u,\ d}\\0&b^{'}_{u,\ d}&y^2_{u,\ d}\end{array}\right).\label{massfinal}
\end{equation}
These real mass matrices can be diagonalized by the following orthogonal transformations:
\begin{eqnarray}
O_{u,~d}^T M_{u,~d}M^T_{u,~d} O_{u, ~d}&=&\left(\begin{array}{ccc}m_{u,~d}^2&0&0\\0&m_{c, ~s}^2&0\\0&0&m_{t, ~b}^2\end{array}\right),\nonumber\\
O_{u^c,~d^c}^T M^T_{u,~d}M_{u^c,~d^c} O_{u^c, ~d^c}&=&\left(\begin{array}{ccc}m_{u,~d}^2&0&0\\0&m_{c, ~s}^2&0\\0&0&m_{t, ~b}^2\end{array}\right).
\end{eqnarray}
The CKM matrix $V_{\rm CKM}$ is then given by
\begin{eqnarray}
V_{\rm CKM}=O_u^T P_q O_d,
\end{eqnarray}
where
\begin{equation}
P_q=P_u^{\dag}P_d=\left(\begin{array}{ccc}1&0&0\\0&e^{i 2\theta_q}&0\\0&0&e^{i\theta_q}\end{array}\right)
\end{equation}
with $\theta_q=\Delta\theta_d-\Delta\theta_u$.

Now it is clear how the $Q_6$ setup reduces the number of parameters in the quark sector.  The total number
of  parameters in the quark sector is nine (four real parameters each in $M_u$ and $M_d$, plus a single phase $\theta_q$),
which should fit ten observables.  Spontaneous CP violation is crucial for this reduction of parameters.
With explicit CP violation, there would have been one more phase parameter.  The single prediction of this
model was numerically studied in Ref. \cite{babu}, and shown to be fully consistent with data.
Here we present a numerical fit to all the quark sector observables, which deviates somewhat from the fit given in
Ref. \cite{babu}.  The difference arises since here we have attempted to be consistent with the recent lattice
determination of light quark masses.  An excellent fit to the quark masses and mixings, including CKM
CP violation, is obtained with the following choice of parameters at a momentum scale of $\mu = 1$ TeV.
$$
m_t^0 = 150.7 ~{\rm GeV},~m_b^0 = 2.5515 ~{\rm GeV}, \hspace{0.5cm}
\theta_q=\Delta\theta_d-\Delta\theta_u=-1.40,
$$
$$ q_u=1.5142 \cdot 10^{-4}, \hspace{0.5cm} b_u=0.0395, \hspace{0.5cm} b^{'}_u=0.0770474, \hspace{0.5cm} y_u=0.99746, $$
\begin{equation} \label{fit}
q_d=0.0043435, \hspace{0.5cm} b_d=0.02609, \hspace{0.5cm}
b^{'}_d=0.69138, \hspace{0.5cm} y_d=0.8100,~\end{equation}
This choice yields at $\mu = 1$ TeV, the following masses and mixings for the quarks:
\begin{eqnarray}
m_u &=& 1.13~{\rm MeV},~~m_c = 0.461~{\rm GeV},~~m_t = 150.50~{\rm GeV}, \nonumber \\
m_d &=& 2.53~{\rm MeV}, ~~m_s = 50.99 {\rm MeV},~~m_b = 2.43~{\rm GeV}, \nonumber \\[0.15in]
|V_{CKM}| &=& \left(\matrix{0.9745 & 0.2244 & 0.0033 \cr
0.2242 & 0.9737 & 0.0408 \cr
0.0093 & 0.0399 & 0.9991} \right) ~,\nonumber \\[0.15in]
 \eta_W &=& 0.3465,
\end{eqnarray}
where $\eta_W$ is the CP violation parameter in the Wolfenstein parametrization.
These values, when extrapolated to lower energy scales, give extremely good agreement with
data \cite{xing}.

We have computed the orthogonal matrices that diagonalize $M_u$ and $M_d$.  These rotation
matrices will be relevant for our discussion of Higgs--induced flavor violation, as well as
FCNC arising via SUSY loop diagrams.  We find
\begin{eqnarray}
O_d=\left(\begin{array}{ccc}0.9840&-0.1782&0.0041\\0.1781&0.9838&0.0188\\-0.0074&-0.0178&0.9998\end{array}\right),~~~
O_{d^c}=\left(\begin{array}{ccc}0.9645&-0.2640&-0.0001\\-0.1817&0.6642&0.7251\\0.1915&-0.6994&0.6886\end{array}\right),\nonumber\\
O_u=\left(\begin{array}{ccc}0.9988&-0.0495&1.17\cdot 10^{-5}\\0.0494&0.9980&0.0395\\-0.0020&-0.0394&0.9992\end{array}\right),~~~ ~~~
O_{u^c}=\left(\begin{array}{ccc}0.9988&0.0496&-6.00\cdot 10^{-6}\\-0.0494&0.9958&0.0771\\0.0038&-0.0770&0.9970\end{array}\right).
\end{eqnarray}

In the case of charged leptons, there is some arbitrariness in the values of $(A,~B,~B',~C)_\ell$ of Eq. (\ref{massmatrix}), since we have three observables (charged lepton masses) and four parameters (without including
the neutrino sector).  We shall present a fit with a simplifying assumption $B'_\ell = B_\ell$.  At $\mu= 1$ TeV,
a consistent fit for all the lepton masses is found with the following input values:
\begin{equation}
A_\ell = 1.67536 ~{\rm GeV},~~B_\ell = B'_\ell = 0.430588 ~{\rm GeV},~~C_\ell = 0.00742877 ~{\rm GeV}~.
\end{equation}
These yield the following eigenvalues at $\mu = 1$ TeV:
\begin{equation}
m_e = 0.4963 ~{\rm MeV}, ~~m_\mu = 104.686~{\rm MeV}, ~~m_\tau = 1779.5~{\rm MeV}.
\end{equation}
These values correspond to the central values of charged lepton masses when extrapolated down
to their respective mass scales \cite{xing}.  The orthogonal matrix that diagonalizes $M_e$ is given
by
\begin{eqnarray}
O_e  = \left(\matrix{0.9976 & 0.0688 & 9.81 \cdot 10^{-4} \cr 0.0664 & -0.9697 & 0.2352 \cr
-0.0171 & 0.2346 & 0.9720}\right)~,
\end{eqnarray}
with $O_{e^c}$ obtained from the above by flipping the signs in the first row and column.

\section{Symmetry breaking and the Higgs boson spectrum}

We now turn to the discussion of symmetry breaking and the Higgs boson spectrum in the model.
We shall confine here to the case of having three pairs of Higgs doublets, and no Higgs singlets
in the low energy theory.
It is however, assumed that singlet fields are present in the full theory, so that
spontaneous $Q_6$ breaking in the singlet sector appears as soft breaking in the doublet
sector. As shown in Ref. \cite{babu}, it is possible to realize such a scenario
while preserving the $1 \leftrightarrow 2$ interchange symmetry for members $(1,~2)$ inside
$Q_6$ doublets.  We seek a consistent picture where CP violating phases are generated in
the Higgs doublet VEVs.  As it turns out, CP  also  has to be softly broken in the
bilinear soft SUSY breaking terms, or else there would be no CP phases in the VEVs.

The superpotential that we consider is the most general one consistent with softly broken
$Q_6$ symmetry, but preserving the $S_2$ interchange symmetry:
\begin{eqnarray}
W_{\rm eff} &=& \mu_1(H^u_1H^d_1+H^u_2H^d_2)+\mu_3H^u_3H^d_3+\mu_{13}(H^u_1+H^u_2)H^d_3{} \nonumber\\
 && {}+\mu_{31}H^u_3(H^d_1+H^d_2)
+\mu_{12}(H^u_1H^d_2+H^d_1H^u_2).
\end{eqnarray}
As mentioned earlier, we make a $45^{\circ}$ rotations in $H^d_1$, $H^d_2$ and  $H^u_1$, $H^u_2$ space, with $\hat{H}^u_{1,~2}=\frac{H^u_1\pm H^u_2}{\sqrt{2}}$
and $\hat{H}^d_{1,~2}=\frac{H^d_1\pm H^d_2}{\sqrt{2}}$, so that the superpotential becomes
\begin{eqnarray}
W_{{\rm eff}} =(\mu_1+\mu_{12})\hat{H}^u_1 \hat{H}^d_1+(\mu_1-\mu_{12})\hat{H}^u_2 \hat{H}^d_2+\mu_3 \hat{H}^u_3 \hat{H}^d_3+\sqrt{2}\mu_{13}\hat{H}^u_1 \hat{H}^d_3+\sqrt{2}\mu_{31}\hat{H}^u_3 \hat{H}^d_1~.
\end{eqnarray}
The redefined fields have $\langle\hat{H}^u_2\rangle=\langle\hat{H}^d_2\rangle=0$.
We work in the hatted basis from now on, and drop the hat on the new fields.

The soft SUSY breaking Lagrangian is given, in the rotated basis, as
\begin{eqnarray}
&&\lefteqn{V_{\rm soft}=(b_1+b_{12})H^u_1 \epsilon H^d_1+(b_1-b_{12})H^u_2 \epsilon H^d_2+b_3 H^u_3\epsilon H^d_3}\hspace{15 cm}\nonumber\\
&&\hspace{1 cm}\lefteqn{+\sqrt{2}b_{13}H^u_1\epsilon H^d_3+ \sqrt{2}b_{31}H^u_3 \epsilon H^d_1+{\rm h. c.}}\nonumber\\
&&\hspace{1 cm}\lefteqn{+m^2_{d1}(|H^d_1|^2+|H^d_2|^2)+m^2_{d3}|H^d_3|^2+m^2_{u1}(|H^u_1|^2+|H^u_2|^2)+m^2_{u3}|H^u_3|^2,}
\end{eqnarray}
where $\epsilon=i\sigma_2$.

The full scalar potential including the soft terms, the $F$ terms and the $D$ terms has the form
\begin{eqnarray}
V&=&M_{d1}^2(|{H^d_1}^0|^2+|{H^d_1}^-|^2)+M_{d3}^2(|{H^d_3}^0|^2+|{H^d_3}^-|^2)\nonumber\\
&&~~+M_{u1}^2(|{H^u_1}^0|^2+|{H^u_1}^+|^2)+M_{u3}^2(|{H^u_3}^0|^2+|{H^u_3}^+|^2)\nonumber\\
 &&~~+\{M_{13}^2( {H^{d}_1}^{0\ast} {H^d_3}^0+ {H^{d}_1}^{-\ast} {H^d_3}^-)
 +M_{31}^2( {H^{u}_3}^{0\ast} {H^u_1}^0+ {H^{u}_3}^{+\ast} {H^u_1}^+)+{\rm h. c.} \}\nonumber\\
 &&~~+M_{d2}^2(|{H^d_2}^0|^2+|{H^d_2}^-|^2)+M_{u2}^2(|{H^u_2}^0|^2+|{H^u_2}^+|^2)\nonumber\\
 &&+\{{b_1'} ({H^u_1}^+ {H^d_1}^--{H^u_1}^0 {H^d_1}^0)+b_3 ({H^u_3}^+ {H^d_3}^--{H^u_3}^0 {H^d_3}^0) \nonumber\\
 &&~~+\sqrt{2}b_{13}({H^u_1}^+ {H^d_3}^--{H^u_1}^0 {H^d_3}^0)+ \sqrt{2}b_{31}({H^u_3}^+ {H^d_1}^--{H^u_3}^0 {H^d_1}^0)\nonumber\\
 &&~~+ {b_2'} ({H^u_2}^+ {H^d_2}^--{H^u_2}^0 {H^d_2}^0)+{\rm h. c.}\}\nonumber\\
&&+\frac{1}{8}(g_1^2+g_2^2)({|H^u_1}^+|^2+{|H^u_1}^0|^2+{|H^u_3}^+|^2+{|H^u_3}^0|^2+{|H^u_2}^+|^2+{|H^u_2}^0|^2 \nonumber\\
&&~~~~~~~~~~~~~~~-{|H^d_1}^-|^2-{|H^d_1}^0|^2-{|H^d_3}^-|^2-{|H^d_3}^0|^2-{|H^d_2}^-|^2-{|H^d_2}^0|^2)^2 \nonumber\\
&&+\frac{1}{2}g_2^2|{H^u_1}^{+}{H^d_1}^{0\ast}+{H^u_1}^{0}{H^d_1}^{-\ast}
+{H^u_3}^{+}{H^d_3}^{0\ast}+{H^u_3}^{0}{H^d_3}^{-\ast}\nonumber\\
+{H^u_2}^{+}{H^d_2}^{0\ast}+{H^u_2}^{0}{H^d_2}^{-\ast}|^2. \hspace{-12cm}
\end{eqnarray}
Here we have redefined new effective parameters for convenience as
\begin{eqnarray}
M_{d1}^2=|\mu_1+\mu_{12}|^2+2 |\mu_{31}|^2+m_{d1}^2,&& ~~~~M_{d3}^2=|\mu_3|^2+2 |\mu_{13}|^2+m_{d3}^2,\nonumber\\
M_{u1}^2=|\mu_1+\mu_{12}|^2+2 |\mu_{13}|^2+m_{u1}^2, &&~~~~M_{u3}^2=|\mu_3|^2+2 |\mu_{31}|^2+m_{u3}^2,\nonumber\\
M_{d2}^2=|\mu_1-\mu_{12}|^2+m_{d1}^2, &&~~~~M_{u2}^2=|\mu_1-\mu_{12}|^2+m_{u1}^2,\nonumber\\
M_{13}^2=\sqrt{2}(\mu_1+\mu_{12})^{\ast}\mu_{13}+\sqrt{2}\mu_3\mu_{31}^{\ast},&&~~~~
M_{31}^2=\sqrt{2}(\mu_1+\mu_{12})\mu_{31}^{\ast}+\sqrt{2}\mu_3^{\ast}\mu_{13},\nonumber\\
b'_1=b_1+b_{12}, ~~~~~~~~~~b'_2=b_1-b_{12}.\hspace{-3.7 cm}
\end{eqnarray}

Before analyzing the spectrum, let us note that the potential should be bounded from below along all
$D$--flat directions.  The following conditions should be satisfied:
\begin{center}
$M_{d1}^2+M_{u1}^2-2|b_1^{'}|>0, ~~~M_{d1}^2+M_{u2}^2>0, ~~~M_{d1}^2+M_{u3}^2-2\sqrt{2}|b_{31}|>0$
\end{center}

$\hspace{3cm}M_{d2}^2+M_{u1}^2>0,~~~M_{d2}^2+M_{u2}^2-2|b_2^{'}|>0,~~~M_{d2}^2+M_{u3}^2>0,$
\begin{eqnarray}\label{bounded}
&&M_{d3}^2+M_{u1}^2-2\sqrt{2}|b_{13}|>0,~~~M_{d3}^2+M_{u2}^2>0,~~~M_{d3}^2+M_{u3}^2-2|b_{3}|>0.
\end{eqnarray}
In our numerical analysis, we shall verify that these conditions are indeed met.

We parametrize the VEVs of the four neutral Higgs fields as
$$
v_{u1}=v \sin \beta \sin{\gamma_u}~e^{i \theta_{u1}}, \hspace{1cm}
v_{u3}=v \sin \beta \cos{\gamma_u}~e^{i \theta_{u3}}, \nonumber {}
$$
\begin{equation}
v_{d1}=v \cos \beta  \sin{\gamma_d}~e^{i \theta_{d1}}, \hspace{1cm}
v_{d3}=v \cos \beta \cos{\gamma_d}~e^{i \theta_{d3}}.
\end{equation}
Thus we have $|v_{u1}|^2+|v_{u3}|^2+|v_{d1}|^2+|v_{d3}|^2 = v^2 = (174~{\rm GeV})^2$.
$\gamma_{u (d)}$ reflect the orientation  of the VEVs in the $H_{u(d)1}- H_{u(d)3}$
space, while $\tan\beta$ is analogous to the up/down VEV ratio of MSSM.

We can rewrite the potential of the $H_1-H_3$ sector of the neutral Higgs fields which acquire VEVs in a compact form:
\begin{eqnarray}
V_N^{(1-3)}&=&\left(\begin{array}{cc}H^{u0\ast}_1&H^{u0\ast}_3\end{array}\right)
\left(\begin{array}{cc}M_{u1}^2&M_{31}^2\\{M_{31}^2}^{\ast}&M_{u3}^2
\end{array}\right)\left(\begin{array}{cc}H^{u0}_1\\H^{u0}_3\end{array}\right)
+\left(\begin{array}{cc}H^{d0\ast}_1&H^{d0\ast}_3\end{array}\right)\left(\begin{array}{cc}M_{d1}^2&M_{13}^2\\{M_{13}^2}^{\ast}&M_{d3}^2
\end{array}\right)\left(\begin{array}{cc}H^{d0}_1\\H^{d0}_3\end{array}\right) \nonumber\\
&&+\left[\left(\begin{array}{cc}H^{u0}_1&H^{u0}_3\end{array}\right)\left(\begin{array}{cc}-b_1^{'}&-\sqrt{2}b_{13}\\-\sqrt{2}b_{31}&-b_3
\end{array}\right)\left(\begin{array}{cc}H^{d0}_1\\H^{d0}_3\end{array}\right)+{\rm h. c.}\right] \nonumber\\
&&+\frac{1}{8}(g_1^2+g_2^2)\left[\left(\begin{array}{cc}H^{u0\ast}_1&H^{u0\ast}_3\end{array}\right)
\left(\begin{array}{cc}H^{u0}_1\\H^{u0}_3\end{array}\right)-\left(\begin{array}{cc}H^{d0\ast}_1&H^{d0\ast}_3\end{array}\right)
\left(\begin{array}{cc}H^{d0}_1\\H^{d0}_3\end{array}\right)\right]^2. \label{higgs}
\end{eqnarray}
This suggests a unitary transformation that would diagonalize the first two matrices in Eq. (\ref{higgs}),
while leaving the $D$--term unaffected.  With such a rotation we have
\begin{eqnarray}
V_N^{(1-3)}&=&\left(\begin{array}{cc}h_1^{\ast}&h_2^{\ast}\end{array}\right)
\left(\begin{array}{cc}m_1^2&0\\0&m_2^2
\end{array}\right)\left(\begin{array}{cc}h_1\\h_2\end{array}\right)
+\left(\begin{array}{cc}h_3^{\ast}&h_4^{\ast}\end{array}\right)
\left(\begin{array}{cc}m_3^2&0\\0&m_4^2
\end{array}\right)\left(\begin{array}{cc}h_3\\h_4\end{array}\right) \nonumber\\
&&+\left[\left(\begin{array}{cc}h_1&h_2\end{array}\right)\left(\begin{array}{cc}m_{13}^2&m_{14}^2\\m_{23}^2&m_{24}^2
\end{array}\right)\left(\begin{array}{cc}h_3\\h_4\end{array}\right)+{\rm h. c.}\right] \nonumber\\
&&+\frac{1}{8}(g_1^2+g_2^2)\left[\left(\begin{array}{cc}h_1^{\ast}&h_2^{\ast}\end{array}\right)
\left(\begin{array}{cc}h_1\\h_2\end{array}\right)-\left(\begin{array}{cc}h_3^{\ast}&h_4^{\ast}\end{array}\right)
\left(\begin{array}{cc}h_3\\h_4\end{array}\right)\right]^2. \label{higgs1}
\end{eqnarray}

The unitary transformations to go from Eq. (\ref{higgs}) to Eq. (\ref{higgs1}) are defined as
\begin{eqnarray}
&&\left(\begin{array}{cc}h_1\\h_2\end{array}\right)=U_U\left(\begin{array}{cc}H^u_1\\H^u_3\end{array}\right)
=Q_u \left(\begin{array}{cc}\cos\omega_u&-\sin\omega_u\\ \sin\omega_u&\cos\omega_u\end{array}\right)
\left(\begin{array}{cc}e^{i\phi_u}&0\\ 0&e^{i(\phi_u+\theta_{M_{31}})}
\end{array}\right)\left(\begin{array}{cc}H^u_1\\H^u_3\end{array}\right), \nonumber\\
&&\left(\begin{array}{cc}h_3\\h_4\end{array}\right)=U_D\left(\begin{array}{cc}H^d_1\\H^d_3\end{array}\right)
=Q_d \left(\begin{array}{cc}\cos\omega_d&-\sin\omega_d\\ \sin\omega_d&\cos\omega_d\end{array}\right)
\left(\begin{array}{cc}e^{i\phi_d}&0\\ 0&e^{i(\phi_d+\theta_{M_{13}})}
\end{array}\right)\left(\begin{array}{cc}H^d_1\\H^d_3\end{array}\right), \label{step1}
\end{eqnarray}
with $\theta_{M_{31}}=\arg(M_{31}^2), ~\theta_{M_{13}}=\arg(M_{13}^2)$ and
\begin{eqnarray}
&&\omega_u=\frac{1}{2}\tan^{-1}\left(\frac{2|M_{31}^2|}{M_{u3}^2-M_{u1}^2}\right),~~~~
\omega_d=\frac{1}{2}\tan^{-1}\left(\frac{2|M_{13}^2|}{M_{d3}^2-M_{d1}^2}\right).
\end{eqnarray}
The two phases $\phi_u$ and $\phi_d$ here are arbitrary. $\phi_u-\phi_d$ does not appear
in the potential (being proportional to $U(1)_Y$ charges). $\phi_u+\phi_d$ can be used to remove one
phase of the bilinear terms in the potential.  $Q_{u,d}$ are arbitrary diagonal phase matrices.
If desired, one can take advantage of these phases to remove all but one phase from the parameters
of the potential.  Since we are interested in going back to the original basis from this rotated
basis, we find it convenient to set $Q_{u,d}$ to be identity.

The other parameters of this transformation are
\begin{eqnarray}
&&m_{1,2}^2=\frac{1}{2}\left[M_{u3}^2+M_{u1}^2\pm\sqrt{(M_{u3}^2-M_{u1}^2)^2+4|M_{31}^2|^2}\right], \nonumber\\
&&m_{3,4}^2=\frac{1}{2}\left[M_{d3}^2+M_{d1}^2\pm\sqrt{(M_{d3}^2-M_{d1}^2)^2+4|M_{13}^2|^2}\right].
\end{eqnarray}
and
\begin{eqnarray}
\left(\begin{array}{cc}m_{13}^2&m_{14}^2\\m_{23}^2&m_{24}^2\end{array}\right)
=U_U^{\ast}\left(\begin{array}{cc}-b_1^{'}&-\sqrt{2}b_{13}\\-\sqrt{2}b_{31}&-b_3\end{array}\right)U_D^{\dag}.
\end{eqnarray}
If we choose
\begin{eqnarray}
&&\lefteqn{\phi_u+\phi_d=\pi+\arg \left[b_1^{'}\sin\omega_u\sin\omega_d
+\sqrt{2}b_{31}\cos\omega_u\sin\omega_d e^{-i\theta_{M_{31}}}\right.} \hspace{12cm}\nonumber\\
&&\lefteqn{\left.\hspace{1.5cm}+\sqrt{2}b_{13}\sin\omega_u\cos\omega_d e^{-i\theta_{M_{13}}}+b_3 \cos\omega_u \cos\omega_d e^{-i(\theta_{M_{31}}+\theta_{M_{13}})}\right],}
\end{eqnarray}
$m_{24}^2$ is real and positive (with $Q_{u,d}$ set to identity). We shall adopt this phase convention
in our numerical study. However, we shall present analytical results that hold in an arbitrary phase
convention.

The task at hand is somewhat simplified, since Eq. (\ref{higgs1}) is relatively simple to analyze.
The eight real neutral Higgs bosons in $H_{1,3}^{u,d}$ can be conveniently parametrized as
\begin{eqnarray}
&&h_1=e^{i\delta_1}\left[v_1+\frac{1}{\sqrt{2}}(\phi_1+ie\phi_5+ia\phi_7+i \frac{v_1}{v}G)\right], \nonumber\\
&&h_2=e^{i\delta_2}\left[v_2+\frac{1}{\sqrt{2}}(\phi_2+if\phi_6+ib\phi_7+i \frac{v_2}{v}G)\right], \nonumber\\
&&h_3=e^{i\delta_3}\left[v_3+\frac{1}{\sqrt{2}}(\phi_3+ig\phi_5+ic\phi_7-i \frac{v_3}{v}G)\right], \nonumber\\
&&h_4=v_4 +\frac{1}{\sqrt{2}}(\phi_4+ih\phi_6+id\phi_7-i\frac{v_4}{v}G).\label{goldstone}
\end{eqnarray}
Here $v_i~ (i=1,~2,~3, ~4)$ are the magnitudes of the  VEVs of the redefined fields $h_i$, and $\delta_i$ are their phases.  Without loss of generality we have taken $v_4$ to be real. $G$ in Eq. (\ref{goldstone}) is the Goldstone field eaten up by the $Z$ gauge boson.  We shall work in the unitary gauge and set $G=0$.  We have checked
explicitly that the $G$ field does not mix with other scalar fields, and that its mass is exactly zero.
The coefficients of various fields in Eq. (\ref{goldstone}) are functions of the $v_i$'s:
\begin{eqnarray}
&&a=\frac{v_1\sqrt{v_2^2+v_4^2}}{v\sqrt{v_1^2+v_3^2}}, ~~~ b=-\frac{v_2\sqrt{v_1^2+v_3^2}}{v\sqrt{v_2^2+v_4^2}},~~~
c=-\frac{v_3\sqrt{v_2^2+v_4^2}}{v\sqrt{v_1^2+v_3^2}},~~~ d=\frac{v_4\sqrt{v_1^2+v_3^2}}{v\sqrt{v_2^2+v_4^2}},\nonumber\\
&&e=\frac{v_3}{\sqrt{v_1^2+v_3^2}}, ~~~~~~ f=\frac{v_4}{\sqrt{v_2^2+v_4^2}},~~~~~~
g=\frac{v_1}{\sqrt{v_1^2+v_3^2}}, ~~~~~~ h=\frac{v_2}{\sqrt{v_2^2+v_4^2}}.
\end{eqnarray}

We shall allow for the soft SUSY breaking parameters $(b_i)$ in the Higgs potential to be complex.
Phase rotations cannot remove all phases from the potential, one phase is unremovable.  Without this phase, the model cannot induce complex VEVs to the doublets, as shown in Ref. \cite{masip} by a geometric argument.  For the case when all parameters in the Higgs potential are
real, we have numerically verified that the CP violating extremum would generate two massless modes, signalling inconsistency with symmetry breaking \cite{masip}.

We take the soft bilinear terms $m_{13}^2, ~m_{14}^2, ~m_{23}^2,~m_{24}^2$ of Eq. (\ref{higgs1})  to be
complex, and denote the phase of $m_{ij}^2$ as $\theta_{ij}$.
The minimization conditions then read as
\begin{eqnarray}
&&m_1^2 v_1+|m_{13}^2| v_3\cos(\theta_{13}+\delta_1+\delta_3)+|m_{14}^2| v_4\cos(\theta_{14}+\delta_1)
+\frac{1}{4}(g_1^2+g_2^2)v_1 (v_1^2+v_2^2-v_3^2-v_4^2)=0, \nonumber\\
&&m_2^2 v_2+|m_{23}^2| v_3\cos(\theta_{23}+\delta_2+\delta_3)+|m_{24}^2| v_4\cos(\delta_2+\theta_{24})
+\frac{1}{4}(g_1^2+g_2^2)v_2 (v_1^2+v_2^2-v_3^2-v_4^2)=0, \nonumber\\
&&m_3^2 v_3+|m_{13}^2| v_1\cos(\theta_{13}+\delta_1+\delta_3)+ |m_{23}^2| v_2\cos(\theta_{23}+\delta_2+\delta_3)
-\frac{1}{4}(g_1^2+g_2^2)v_3 (v_1^2+v_2^2-v_3^2-v_4^2)=0, \nonumber\\
&&m_4^2 v_4+|m_{14}^2| v_1\cos(\theta_{14}+\delta_1)+|m_{24}^2| v_2\cos(\delta_2+\theta_{24})
-\frac{1}{4}(g_1^2+g_2^2)v_4 (v_1^2+v_2^2-v_3^2-v_4^2)=0, \nonumber\\
&&
|m_{13}^2|(v_1^2+v_3^2)\sin(\theta_{13}+\delta_1+\delta_3)+|m_{23}^2|v_1v_2\sin(\theta_{23}+\delta_2+\delta_3)+|m_{14}^2|v_3 v_4 \sin(\theta_{14}+\delta_1) = 0,
\nonumber\\
&&
|m_{24}^2| (v_2^2+v_4^2) \sin(\theta_{24}+\delta_2) + |m_{14}^2| v_1 v_2 \sin(\theta_{14}+\delta_1) + |m_{23}^2| v_3 v_4 \sin(\theta_{23}+\delta_2+\delta_3)=0,
\nonumber\\
&&|m_{14}^2| v_1 v_4 \sin(\theta_{14}+\delta_1)-|m_{23}^2| v_2 v_3 \sin(\theta_{23}+\delta_2+\delta_3)=0.
\end{eqnarray}

Denoting the squared matrix for $\phi_i$, $i=1, 2, \ldots 7$ from the $H_1-H_3$ sector as
\begin{equation}
\mathcal{M}_{0, (1-3)}^2=\mathcal{M}_{ij}^2, \label{mass1}
\end{equation}
we obtain
\begin{eqnarray}
&&\mathcal{M}_{11}^2=\lambda v_1^2+\kappa \frac{v_2 v_4}{v_1^2}[\cot(\theta_{14}+\delta_1)
-\cot(\theta_{13}+\delta_1+\delta_3)],\nonumber\\
&&\mathcal{M}_{22}^2=\lambda v_2^2+\kappa \frac{v_4}{v_2}[\cot(\theta_{23}+\delta_2+\delta_3)
-\cot(\theta_{24}+\delta_2)],\nonumber\\
&&\mathcal{\mathcal{M}}_{33}^2=\lambda v_3^2+\kappa \frac{v_2 v_4}{v_3^2}[\cot(\theta_{23}+\delta_2+\delta_3)
-\cot(\theta_{13}+\delta_1+\delta_3)],\nonumber\\
&&\mathcal{M}_{44}^2=\lambda v_4^2+\kappa \frac{v_2}{v_4}[\cot(\theta_{14}+\delta_1)
-\cot(\theta_{24}+\delta_2)],\nonumber\\
&&\mathcal{M}_{55}^2= \kappa {v_2 v_4 \over v_1^2+v_3^2}[{v_3^2 \over v_1^2} \cot(\theta_{14}+\delta_1) +{v_1^2 \over v_3^2} \cot(\theta_{23}+\delta_2+\delta_3) - {(v_1^2+v_3^2)^2 \over v_1^2v_3^2}\cot(\theta_{13}+\delta_1+\delta_3)],
\nonumber\\
&&\mathcal{M}_{66}^2= \kappa {1 \over v_2v_4( v_2^2+v_4^2)}[v_2^4 \cot(\theta_{14}+\delta_1) +v_4^4 \cot(\theta_{23}+\delta_2+\delta_3) - (v_2^2+v_4^2)^2 \cot(\theta_{24}+\delta_2)],
\nonumber\\
&&\mathcal{M}_{77}^2= \kappa  {v_2v_4(v_1^2+v_2^2+v_3^2+v_4^2) \over (v_1^2+v_3^2)( v_2^2+v_4^2)}
[\cot(\theta_{14}+\delta_1) + \cot(\theta_{23}+\delta_2+\delta_3)],
\nonumber\\
&&\mathcal{M}_{12}^2=\lambda v_1 v_2 ,~~~~~~
\mathcal{M}_{13}^2=-\lambda v_1 v_3+\kappa \frac{v_2 v_4}{v_1 v_3}\cot(\theta_{13}+\delta_1+\delta_3),\nonumber\\
&&\mathcal{M}_{14}^2=-\lambda v_1 v_4-\kappa \frac{v_2}{v_1}\cot(\theta_{14}+\delta_1),~~~~~~
\mathcal{M}_{15}^2=-\kappa \frac{v_2 v_4}{v_3 \sqrt{v_1^2+v_3^2}},~~~~~~
\mathcal{M}_{16}^2=\kappa \frac{v_2^2}{v_1 \sqrt{v_2^2+v_4^2}},\nonumber\\
&&\mathcal{M}_{17}^2=\kappa {v_2 v_4 \over v_1}{\sqrt{v_1^2+v_2^2+v_3^2+v_4^2} \over \sqrt{v_1^2+v_3^2}\sqrt{v_2^2+v_4^2}},~~~~~~
\mathcal{M}_{23}^2=-\lambda v_2 v_3
-\kappa \frac{v_4}{v_3}\cot(\theta_{23}+\delta_2+\delta_3),\nonumber\\
&&\mathcal{M}_{24}^2=-\lambda v_2 v_4+\kappa \cot(\theta_{24}+\delta_2),~~~~~~
\mathcal{M}_{25}^2=\kappa \frac{v_1 v_4}{v_3\sqrt{v_1^2+v_3^2}},~~~~~~
\mathcal{M}_{26}^2=-\kappa {v_2 \over \sqrt{v_2^2+v_4^2}},\nonumber\\
&&\mathcal{M}_{27}^2=-\kappa {v_4 \sqrt{v_1^2+v_2^2+v_3^2+v_4^2} \over \sqrt{v_1^2+v_3^2}\sqrt{v_2^2+v_4^2}},~~~~~~
\mathcal{M}_{34}^2=\lambda v_3 v_4 ,~~~~~~
\mathcal{M}_{35}^2=-\kappa {v_2 v_4 \over v_1 \sqrt{v_1^2+v_3^2}},\nonumber \\
&&\mathcal{M}_{36}^2=\kappa \frac{v_4^2}{v_3\sqrt{v_2^2+v_4^2}},~~~~~
\mathcal{M}_{37}^2=-\kappa \frac{v_2 v_4}{v_3}{\sqrt{v_1^2+v_2^2+v_3^2+v_4^2} \over \sqrt{v_1^2+v_3^2} \sqrt{v_2^2+v_4^2}},~~~~~~
\mathcal{M}_{45}^2=\kappa \frac{v_2 v_3}{v_1 \sqrt{v_1^2+v_3^2}},\nonumber \\
&&\mathcal{M}_{46}^2=-\kappa {v_4 \over \sqrt{v_2^2+v_4^2}},~~~~~~
\mathcal{M}_{47}^2=\kappa {v_2 \sqrt{v_1^2+v_2^2+v_3^2+v_4^2} \over \sqrt{v_1^2+v_3^2}\sqrt{v_2^2+v_4^2}},\nonumber\\
&&\mathcal{M}_{56}^2=\kappa {1 \over v_1 v_3 \sqrt{v_1^2+v_3^2} \sqrt{v_2^2+v_4^2}}[v_2^2v_3^2 \cot(\theta_{14}+\delta_1)+ v_1^2v_4^2 \cot(\theta_{23}+\delta_2+\delta_3)]
,\nonumber\\
&&\mathcal{M}_{57}^2=\kappa {v_2 v_4 \over v_1 v_3} {\sqrt{v_1^2+v_2^2+v_3^2+v_4^2} \over (v_1^2+v_3^2)\sqrt{v_2^2+v_4^2}}[v_3^2 \cot(\theta_{14}+\delta_1)-v_1^2\cot(\theta_{23}+\delta_2+\delta_3)]
,\nonumber\\
&&\mathcal{M}_{67}^2=\kappa {\sqrt{v_1^2+v_2^2+v_3^2+v_4^2} \over \sqrt{v_1^2+v_3^2}(v_2^2+v_4^2)}[v_2^2 \cot(\theta_{14}+\delta_1)-v_4^2\cot(\theta_{23}+\delta_2+\delta_3)].
\end{eqnarray}
Here we have defined $\lambda=(g_1^2+g_2^2)/2=M_Z^2/v^2$ and $\kappa =m_{24}^2\sin(\theta_{24}+\delta_2)$.

The potential of the $H_2^u-H_2^d$ fields which do not acquire VEVs is
\begin{eqnarray}
V_N^{(2)}&=&M_{u2}^2|H_2^u|^2+M_{d2}^2|H_2^d|^2-\{b_2^{'}H_2^u H_2^d+{\rm h. c.}\}\nonumber \\
&+&\frac{g_1^2+g_2^2}{8}(|H_2^u|^2-|H_2^d|^2+|v_{u1}|^2+|v_{u3}|^2-|v_{d1}|^2-|v_{d3}|^2)^2~.
\end{eqnarray}
The corresponding squared mass matrix for the scalars in the basis
$({\rm Re}H^u_2, ~{\rm Im}H^u_2,~{\rm Re}H^d_2,~{\rm Im}H^d_2)$ is
\begin{eqnarray}\label{mass2}
\mathcal{M}_{0(2)}^2=\left(\begin{array}{cccc}M_{u2}^2-{m_Z^2 \over 2} \cos 2\beta&0&{\rm Re} b_2^{'}&-{\rm Im} b_2^{'}\\0&M_{u2}^2-{m_Z^2 \over 2} \cos 2\beta&-{\rm Im} b_2^{'}&-{\rm Re} b_2^{'}\\
{\rm Re} b_2^{'}&-{\rm Im} b_2^{'}&M_{d2}^2+{m_Z^2 \over 2} \cos 2\beta &0\\-{\rm Im} b_2^{'}&-{\rm Re} b_2^{'}&0&M_{d2}^2+{m_Z^2 \over 2}\cos 2\beta\end{array}\right)~.
\end{eqnarray}
This matrix has two pairs of degenerate eigenstates, owing to an unbroken $U(1)$ symmetry.

The $H_1^{u,d}-H_3^{u,d}$ sector charged Higgs boson mass matrix is, in the basis $\{{H^u_1}^+,~{H^u_3}^+,~{{H^d_1}^-}^{\ast},~{{H^d_3}^-}^{\ast}\}$,
\begin{center}
$\mathcal{M}_{\pm(1-3)}^2=(\mathcal{M}^2)_{ij}$,
\end{center}
with
\begin{eqnarray}
&&\mathcal{M}^{2}_{11}=M_{u1}^2-\frac{1}{2}m_Z^2 \cos 2\beta+\frac{1}{2}g_2^2|v_{d1}|^2,~~~~~~
\mathcal{M}^{2}_{22}=M_{u3}^2-\frac{1}{2}m_Z^2 \cos 2\beta+\frac{1}{2}g_2^2|v_{d3}|^2, \nonumber\\
&&\mathcal{M}^{2}_{33}=M_{d1}^2+\frac{1}{2}m_Z^2 \cos 2\beta+\frac{1}{2}g_2^2|v_{u1}|^2,~~~~~~
\mathcal{M}^{2}_{44}=M_{d3}^2+\frac{1}{2}m_Z^2 \cos 2\beta+\frac{1}{2}g_2^2|v_{u3}|^2, \nonumber\\
&&\mathcal{M}^{2}_{12}={\mathcal{M}^{2}_{21}}^{\ast}=M_{31}^2+\frac{1}{2}g_2^2 v_{d1}^{\ast}v_{d3},~~~~~~~~~~
\mathcal{M}^{2}_{13}={\mathcal{M}^{2}_{31}}^{\ast}=b_1'+\frac{1}{2}g_2^2 v_{u1}^{\ast}v_{d1}^{\ast}, \nonumber\\
&&\mathcal{M}^{2}_{14}={\mathcal{M}^{2}_{41}}^{\ast}=\sqrt{2}b_{13}+\frac{1}{2}g_2^2v_{u3}^{\ast}v_{d1}^{\ast},
~~~~~~~~~\mathcal{M}^{2}_{23}={\mathcal{M}^{2}_{32}}^{\ast}=\sqrt{2}b_{31}+\frac{1}{2}g_2^2v_{u1}^{\ast}v_{d3}^{\ast},
\nonumber\\
&&\mathcal{M}^{2}_{24}={\mathcal{M}^{2}_{42}}^{\ast}=b_{3}+\frac{1}{2}g_2^2v_{u3}^{\ast}v_{d3}^{\ast},~~~~~~~~~~~~~~
\mathcal{M}^{2}_{34}={\mathcal{M}^{2}_{43}}^{\ast}=M_{13}^2+\frac{1}{2}g_2^2 v_{u1}v_{u3}^{\ast}.
\end{eqnarray}

Finally, the $H_2^u-H_2^d$ sector charged Higgs mass matrix is, in the basis $\{{H^u_2}^+,~{{H^d_2}^-}^{\ast}\}$,
\begin{eqnarray}\label{mass3}
\mathcal{M}^2_{\pm(2)}=\left(\begin{array}{cc}M_{u2}^2-\frac{1}{2}m_Z^2 \cos 2\beta&{b_2^{'}}\\{b_2^{'}}^{\ast}&M_{d2}^2+\frac{1}{2}m_Z^2 \cos 2\beta\end{array}\right)
\end{eqnarray}

Now we present two sets of numerical fits (cases (1) and (2)) which show the consistency of symmetry breaking.  We are interested in choosing the SUSY breaking parameters (including the $\mu$ terms) around the TeV scale, guided by
arguments of naturalness.  At the same time we wish the spectrum to be consistent with FCNC constraints
arising from meson--antimeson mixings.  We have explored parameter space of the Higgs potential where both these
constraints are met.  For the FCNC constraint, we allow the new Higgs exchange contribution to $\Delta M$
be not more than the experimentally measured values.

\noindent{\bf Case (1)}

The parameters in the original Higgs potential of Eq. (\ref{higgs}) are taken to have the following values.
\begin{eqnarray}
&&M_{d1}=3.754 {~\rm TeV},~~  M_{d3}=3.586 {~\rm TeV},~~ M_{u1}=4.782 {~\rm TeV}, ~~M_{u3}=2.152 {~\rm TeV},\nonumber\\
&&M_{31}=2.336~ e^{i 0.792} {~\rm TeV},~~M_{13}=1.346 ~e^{-i 1.205} {~\rm TeV}, ~~b_1^{'}=3.144~e^{i 2.963}{~\rm TeV}^2,  \nonumber\\
&&b_3=3.196 ~e^{i 2.064}{~\rm TeV}^2,~~b_{31}=4.052~ e^{i 2.186}{~\rm TeV}^2, ~~b_{13}=3.438~ e^{i 3.109}{~\rm TeV}^2, \nonumber\\
&&M_{u2}=4.550{~\rm TeV}, ~~~~~~~~~~~M_{d2}=4.850 {~\rm TeV} ~~~~~~~~~~~b_2^{'}=0.000 {~\rm TeV}^2,
\end{eqnarray}
In the representation of Eq. (\ref{higgs1}) this choice corresponds to
\begin{eqnarray}
&&m_1=4.937 {~\rm TeV},~~  m_2=1.767 {~\rm TeV},~~ m_3=3.923 {~\rm TeV}, m_4=3.401 {~\rm TeV},\nonumber\\
&&m_{13}=1.851 e^{-i 1.437} {~\rm TeV},~~m_{14}=2.736 e^{-i 0.732} {~\rm TeV}, ~~m_{23}=2.442~e^{i 1.347}{~\rm TeV},  \nonumber\\
&&m_{24}=2.104 {~\rm TeV}~.
\end{eqnarray}
For completeness we also give values of other parameters,  $\omega_u=0.70, ~~\omega_d=0.622, ~~\phi_u+\phi_d=1.005$.

We obtain numerically the VEV parameters to be
\begin{equation}
\tan \beta=2.00, ~~\Delta\theta_d=-0.03,~~\Delta\theta_u=1.37, ~~\tan\gamma_d=2.50, ~~\tan\gamma_u=0.33.
\end{equation}

The mass eigenvalues of the Higgs bosons in the $H_1-H_3$  sector are found to be
\begin{eqnarray}
M_{h0}=(99.4, ~115.1) {~\rm GeV},  ~~M_1=3.299 {~\rm TeV},~~M_2-M_1=0.226 {\rm ~GeV}, \hspace{1 cm}\nonumber\\
M_3=4.161{~\rm TeV}, ~~M_4-M_3=0.411 {\rm ~GeV}, ~~M_5=5.124 {~\rm TeV},~~M_6-M_5=0.040 {\rm ~GeV}.\label{spectrum1}
\end{eqnarray}
Note the appearance of nearly degenerate states $(M_1,~M_2)$ etc, with their mass splitting being proportional
to $m_Z^2/4$. The Higgs bosons from the $H_2$ sector have degenerate masses given by
\begin{eqnarray}
M_7=M_8=4.850{~\rm TeV}~~~~~M_9=M_{10}= 4.550 {~\rm TeV}.\label{deg}
\end{eqnarray}

The charged Higgs bosons are nearly degenerate with its neutral partner, so we list the mass splittings:
\begin{eqnarray}
M_{\pm1}-M_1=-0.532 {\rm ~GeV}, ~~~M_{\pm2}-M_3=-0.156{\rm ~GeV},~~~M_{\pm3}-M_5=0.032{\rm ~GeV}~.
\end{eqnarray}

In the $(H_2^u-H_2^d)$  sector, the two charged Higgs bosons are degenerate with the neutral ones given
in Eq. (\ref{deg}).

The mass eigenstates $H_i$ are mixtures of $h_i, ~i=1,~2,~\dots~7$ states in the (1-3) sector. The orthogonal transformation that
diagonalizes the mass matrix of Eq. (\ref{mass1}) is
\begin{eqnarray}
H^k=\left(\begin{array}{ccccccc}
0.0662&0.8919&0.2708&0.3562&8.60\cdot 10^{-7}&1.23\cdot 10^{-6}&2.15\cdot 10^{-6}\\
0.0314&-0.0023&0.0427&-0.0324&-0.4002&0.8800&0.2482\\
0.3322&-0.2620&-0.3269&0.8428&0.0204&0.0293&0.0509\\
-0.0357&0.0026&-0.0484&0.0368&0.1514&0.3354&-0.9272\\
-0.0644&0.3645&-0.9010&-0.2159&0.0231&0.0332&0.0578\\
0.0430 &-0.0032&0.0584&-0.0444&0.9029&0.3311&0.2607\\
-0.9365&-0.0553&-0.0289&0.3345&0.0279&0.0401&0.0697
\end{array}\right)\left(\begin{array}{ccccccc}
h_1\\h_2\\h_3\\h_4\\h_5\\h_6\\h_7\end{array}\right),\hspace{0.5cm}\nonumber
\end{eqnarray}
\begin{eqnarray}
\label{masseigen1}
\end{eqnarray}
with $k=0, \cdots 6$. Since $b_2' =0$ in this case, the $H_2^0$ mass matrix is diagonal, and thus
the mass eigenstates are the original state.

\noindent{\bf Case (2)}

Here we take the input parameters corresponding to Eq. (\ref{higgs}) to be
\begin{eqnarray}
&&M_{d1}=3.980 {~\rm TeV},~~  M_{d3}=5.412 {~\rm TeV},~~ M_{u1}=2.765 {~\rm TeV}, ~~M_{u3}=3.692 {~\rm TeV},\nonumber\\
&&M_{31}=2.825~ e^{i 0.781} {~\rm TeV},~~M_{13}=1.693 ~e^{-i 0.949} {~\rm TeV}, ~~b_1^{'}=3.698 ~e^{i 1.495}{~\rm TeV}^2,  \nonumber\\
&&b_3=3.097 ~e^{i 1.522}{~\rm TeV}^2,~~b_{31}=7.420~ e^{i 2.428}{~\rm TeV}^2, ~~b_{13}=1.840~ e^{-i 2.772}{~\rm TeV}^2,\nonumber\\
&&M_{u2}=3.550{~\rm TeV},~~~~~~~~~~ M_{d2}=5.850 {~\rm TeV}, ~~~~~~~~~~~b_2^{'}=1.234 e^{i1.56}{~\rm TeV}^2.
\end{eqnarray}
This choice corresponds to parameters in Eq. (\ref{higgs1}) to be
\begin{eqnarray}
&&m_1=4.377 {~\rm TeV},~~  m_2=1.154 {~\rm TeV},~~ m_3=5.466 {~\rm TeV}, m_4=3.906 {~\rm TeV},\nonumber\\
&&m_{13}=3.281 e^{i 1.271} {~\rm TeV},~~m_{14}=1.702 e^{i 0.974} {~\rm TeV}, ~~m_{23}=3.190~e^{-i 0.501}{~\rm TeV},  \nonumber\\
&&m_{24}=2.326 {~\rm TeV},
\end{eqnarray}
with $\omega_u=-0.501, ~~\omega_d=-0.606, ~~\phi_u+\phi_d=4.786.$

The Higgs VEV parameters are found for this input to be
\begin{equation}
\tan \beta=2.40, ~~\Delta\theta_d=-0.06, ~~\Delta\theta_u=1.34, ~~\tan\gamma_d=1.80, ~~\tan\gamma_u=1.00.
\end{equation}

The mass spectrum of Higgs boson in the $H_1-H_3$ sector is
\begin{eqnarray}
M_{h0}=(104.1, ~119.2) {~\rm GeV},  ~~M_1=2.869 {~\rm TeV}, ~~M_2-M_1= 0.325 {\rm ~GeV}\hspace{1cm}\nonumber\\
M_3=5.114 {~\rm TeV},~~M_4-M_3=0.132 {\rm ~GeV}, ~~M_5=5.658 {~\rm TeV}, ~~M_6-M_5=0.087 {\rm ~GeV},\label{spectrum2}
\end{eqnarray}
while the mass eigenvalues of Eq. (\ref{mass2}) are
\begin{eqnarray}\label{deg1}
M_7=M_8=5.856{\rm ~TeV}~~~~~M_9=M_{10}= 3.541 {~\rm TeV}.
\end{eqnarray}

The charged Higgs boson masses are given by
\begin{eqnarray}
M_{\pm1}-M_1=0.225 {\rm ~GeV}, ~~~M_{\pm2}-M_3=0.182{\rm ~GeV},~~~M_{\pm3}-M_5=-0.064{\rm ~GeV}~.
\end{eqnarray}
with the remaining two charged Higgs bosons being degenerate with the neutral ones given in Eq. (\ref{deg1}).

The orthogonal matrix that diagonalizes Eq. (\ref{mass1}) is
\begin{eqnarray}
H^k=\left(\begin{array}{ccccccc}
0.3919&0.8356&0.2620&0.2819&1.02\cdot 10^{-4}&-5.99 \cdot10^{-5}&7.35\cdot 10^{-5}\\
0.4761&-0.2234&-0.1898&0.1764&0.6693&0.4070&0.2065\\
-0.5233&-0.0380&0.4553&0.4166&0.4223&-0.2386&0.3295\\
0.5065&-0.2376&0.1014&-0.0942&0.0903&-0.8105&0.0535\\
-0.0893&0.2828&0.0490&-0.7599&0.1969&0.0134&0.5415\\
-0.0625&0.0293&0.3803&-0.3534&0.4512&0.0437&-0.7212\\
0.2783&-0.3363&0.7285&-0.0671&-0.3510&0.3422&0.1805
\end{array}\right)\left(\begin{array}{ccccccc}
h_1\\h_2\\h_3\\h_4\\h_5\\h_6\\h_7\end{array}\right),\hspace{0.5cm}\label{masseigen2}
\end{eqnarray}
The matrix diagonalizing  Eq. (\ref{mass2}) is
\begin{eqnarray}
H^k=\left(\begin{array}{cccc}-0.0568&0.000&0.9984&0.0000\\0.0000&-0.0568&0.0000&0.9984\\
0.0000&0.9984&0.0000&0.0568\\0.9984&0.0000&0.0568&0.0000
\end{array}\right)\left(\begin{array}{cccc}
{\rm Re}(H^u_2)\\{\rm Im}(H^u_2)\\{\rm Re}(H^d_2)\\{\rm Im}(H^d_2)\end{array}\right),
\end{eqnarray}
with $k=7, \cdots 10$.

In these fits, $M_{h_0}$ is the light standard model--like Higgs boson mass, for which radiative corrections
are significant.  In our computation we have included known two loop corrections. The two values listed
for $M_{h_0}$ correspond to zero and maximal left--right stop mixing ($X_t = 0$ or 6).  We have taken $m_t=174{~\rm GeV}$,
$M_{\rm SUSY}=1.5 {~\rm TeV}$ and $\alpha_s(m_t)=0.108$ for these evaluations and used the analytic approximation
given in Ref. \cite{carena}.

An interesting feature of these two fits is that the diagonal entries of the quadratic
mass matrix of the potential  of Eq. (\ref{higgs}) are all positive.  This of course does not preclude some
soft squared masses turning negative as in the MSSM via large top quark Yukawa coupling (since the diagonal
entries also receive $\mu$ term contributions), however, this is
not necessary for symmetry breaking to be triggered.   Yet, one of the eigenvalues of this matrix is
negative, which facilitates symmetry breaking.  For the two cases we find these eigenvalues to be
\begin{eqnarray}
&~&{\rm Case} ~{\bf (1)}: \{(5.123~{\rm TeV})^2,~~~ (4.161~{\rm TeV})^2,~~~     (3.300~{\rm TeV})^2,~~~-(38.682~{\rm GeV})^2  \}~,
\nonumber \\
&~&{\rm Case}~ {\bf (2)}: \{(5.658~{\rm TeV})^2,~~~     (5.115~{\rm TeV})^2,~~~     (2.869~{\rm TeV})^2,~~~    -(45.40~{\rm GeV})^2\}~.
\end{eqnarray}
The conditions for boundedness of the potential listed in Eq. (\ref{bounded}) are found to be satisfied
for both cases.

\subsection{Neutralino and Chargino masses}

The symmetry breaking parameters do not fully determine the masses of the neutralinos and the charginos.
Here we present analytical results for their mass matrices.

The mass matrix of $\tilde{H}_1-\tilde{H}_3$ sector neutralino in the basis of $\{\widetilde{B},~\widetilde{W}^0, ~\widetilde{{H^u_1}}^0, ~\widetilde{{H^u_3}}^0,~\widetilde{{H^d_1}}^0, ~\widetilde{{H^d_3}}^0\}$ is
\begin{eqnarray}
\mathcal{M}_{\chi^0(13)}=\left(\begin{array}{cccccc}M_{\tilde{B}}&0&\frac{g_2v_{u1}}{\sqrt{2}}&\frac{g_2v_{u3}}{\sqrt{2}}&-\frac{g_2v_{d1}}{\sqrt{2}}&-\frac{g_2v_{d3}}{\sqrt{2}}\\
0&M_{\tilde{W}}&-\frac{g_1v_{u1}}{\sqrt{2}}&-\frac{g_1v_{u3}}{\sqrt{2}}&\frac{g_1v_{d1}}{\sqrt{2}}&\frac{g_1v_{d3}}{\sqrt{2}}\\
\frac{g_2v_{u1}}{\sqrt{2}}&-\frac{g_1v_{u1}}{\sqrt{2}}&0&0&-(\mu_1+\mu_{12})&-\sqrt{2}\mu_{13}\\
\frac{g_2v_{u3}}{\sqrt{2}}&-\frac{g_1v_{u3}}{\sqrt{2}}&0&0&-\sqrt{2}\mu_{31}&-\mu_3\\
-\frac{g_2v_{d1}}{\sqrt{2}}&\frac{g_1v_{d1}}{\sqrt{2}}&-(\mu_1+\mu_{12})&-\sqrt{2}\mu_{31}&0&0\\
-\frac{g_2v_{d3}}{\sqrt{2}}&\frac{g_1v_{d3}}{\sqrt{2}}&-\sqrt{2}\mu_{13}&-\mu_3&0&0\end{array}\right)~.
\end{eqnarray}
The mass matrix of the $\tilde{H}_2$ sector in the basis   $\{~\widetilde{{H^u_2}}^0 ,~\widetilde{{H^d_2}}^0\}$ is
\begin{eqnarray}
\mathcal{M}_{\chi^0(2)}\left(\begin{array}{cc}0&-(\mu_1-\mu_{12})\\-(\mu_1-\mu_{12})&0\end{array}\right)~.
\end{eqnarray}

The mass matrix of charginos of the $\tilde{H}_1-\tilde{H}_3$ sector in the basis $\{\widetilde{W}^+, ~\widetilde{{H^u_1}}^+, ~\widetilde{{H^u_3}}^+ ,~\widetilde{W}^-, ~\widetilde{{H^d_1}}^-, ~\widetilde{{H^d_3}}^-\}$ has a block--diagonal form:
\begin{eqnarray}
\mathcal{M}_{\chi^{\pm}(13)}=\left(\begin{array}{cc}0& \bf{X}^T\\ \bf{X}&0\end{array}\right)
\end{eqnarray}
with
\begin{eqnarray}
\bf{X}=\left(\begin{array}{ccc}M_{\tilde{W}}&g_1 v_{u1}&g_1 v_{u3}\\g_1 v_{d1}&\mu_1+\mu_{12}&\sqrt{2}\mu_{31}\\
g_1 v_{d3}&\sqrt{2}\mu_{13}&\mu_3\end{array}\right)~.
\end{eqnarray}
The chargino mass matrix in the $\tilde{H}^u_2-\tilde{H}^d_2$ sector in the basis of $\{\widetilde{{H^u_2}}^+,~\widetilde{{H^d_2}}^-\}$ is
\begin{eqnarray}
\mathcal{M}_{\chi^{\pm}(2)}=\left(\begin{array}{cc}0&\mu_1-\mu_{12}\\\mu_1-\mu_{12}&0\end{array}\right)
\end{eqnarray}

\section{Tree level Higgs induced FCNC processes}

In this section we discuss various FCNC processes mediated by tree--level neutral
Higgs boson exchange.

\subsection{Neutral meson mixing via Higgs exchange}

Accurate measurements exist \cite{pdg} for neutral meson--antimeson mixings in the $K^0-\overline{K^0}$, $B_d^0-\overline{B_d^0}$, $B_s^0-\overline{B_s^0}$ and in $D^0-\overline{D^0}$ sectors.
In the $Q_6$ model there are new contributions to these mixings arising through tree--level Higgs exchange.
These new contributions
will modify the SM predictions, which are all in good agreement with data.
Here we compute these new contributions, following the analysis of Ref. \cite{he}, with updated QCD corrections and hadronic matrix elements.

The Yukawa coupling  $\alpha_{u,\ d},\ \beta_{u,\ d}, \ \beta^{'}_{u,\ d},\ \delta_{u,\ d}$ of
Eq. (\ref{Md}) can be determined from the mass matrix Eq. (\ref{massfinal}):
$$ \alpha_{u,\ d}=\frac{m^0_{t,\ b}y_{u,\ d}^2}{|v_{u,~d3}|}, \hspace{1cm}
\beta_{u,\ d}=\frac{m^0_{t,\ b}b_{u,\ d}}{|v_{u, ~d1}|} $$

\begin{equation}
\beta^{'}_{u,\ d}=\frac{m^0_{t,\ b}b^{'}_{u,\ d}}{|v_{u, ~d1}|}, \hspace{1cm}
\delta_{y,\ d}=\frac{m^0_{t,\ b}q_{u,\ d}/y_{u,\ d}}{|v_{u,~d3}|},
\end{equation}
Using the input values given in Eq. (\ref{fit}) we get for the two cases

\noindent{\bf Case (1)}

$\hspace{2cm}\alpha_d=0.0409,~~~~~~\beta_d=6.51\cdot 10^{-4},~~~~~~\beta_d^{'}=0.0173,~~~~~~\delta_d=3.35\cdot 10^{-4},$\\
$~~~~~~\hspace{2cm}\alpha_u=0.7195,~~~~~~\beta_u=0.0858,~~~~~~\beta_u^{'}=0.1672,~~~~~~\delta_u=1.10\cdot 10^{-4}.$

\noindent{\bf Case (2)}

$\hspace{2cm}\alpha_d=0.0526,~~~~~~\beta_d=7.46\cdot 10^{-4},~~~~~~\beta_d^{'}=0.0198,~~~~~~\delta_d=4.30\cdot 10^{-4},$\\
$~~~~~~\hspace{2cm}\alpha_u=0.9354,~~~~~~\beta_u=0.0372,~~~~~~\beta_u^{'}=0.0724,~~~~~~\delta_u=1.43\cdot 10^{-4}.$

After $45^{\circ}$ rotation in the $Q_6$ doublet space, the Yukawa coupling matrices in the down sector are
\begin{eqnarray}
Y_{{d1}}=O_{d}^T P_{d}\left(\begin{array}{ccc}0&0&0\\0&0&\beta_{d}\\0&\beta_{d}^{'}&0\end{array}\right)P_{d^c}O_{d^c},~~~~
Y_{d2}=O_{d}^T P_{d}\left(\begin{array}{ccc}0&0&\beta_{d}\\0&0&0\\\beta_{d}^{'}&0&0\end{array}\right)P_{d^c}O_{d^c},\nonumber\\
Y_{d3}=O_{d}^T P_{d}\left(\begin{array}{ccc}0&\delta_{d}&0\\-\delta_{d}&0&0\\0&0&\alpha_{d}\end{array}\right)P_{d^c}O_{d^c},
\hspace{3cm}
\end{eqnarray}
where $P_d, ~P_{d^c}$ are defined in Eq. (\ref{p1}). The Yukawa couplings in the up--quark sector and the
charged lepton sector are similar.

The new Higgs--mediated contributions to  $\Delta F=2$ Hamiltonian, responsible for the neutral meson--antimeson mixings has the form \cite{he}
\begin{equation}
H_{\rm eff}=-\frac{1}{2{M_k}^2} \left( \bar{q}_i\left[Y^k_{ij}\frac{1+\gamma_5}{2}+{Y^k_{ji}}^*\frac{1-\gamma_5}{2}\right]q_j\right)^2.
\end{equation}
Here $q_{i,j}$ are the relevant quark fields contained in the meson. $Y^k_{ij}$ are the Yukawa couplings
of $q_i$, $q_j$ with Higgs mass eigenstate $H^k$ mediating FCNC interactions, $k=1,~2,~\ldots 10$ in our model, 6 from the $(H_1-H_3)$ sector and 4 from the $H_2$ sector.
(The light standard model--like Higgs boson has practically no FCNC couplings.)
$Y^k_{ij}$ can be obtained via inverse transformations, Eq. (\ref{step1}),
(\ref{goldstone}) and (\ref{masseigen1}) or (\ref{masseigen2}).

We obtain
\begin{eqnarray}
&&\lefteqn{M^{\phi}_{12}=\langle\phi|H_{{\rm eff}}|\bar{\phi}\rangle
=-\frac{{f_{\phi}}^2m_{\phi}}{2{M_k}^2}\Big[-\frac{5}{24}
\frac{m_{\phi}^2}{(m_{q_i}+m_{q_j})^2}\left({Y^k_{ij}}^2+{{Y^k_{ji}}^*}^2\right)\cdot B_{2}\cdot \eta_2(\mu)}\hspace{14 cm}\nonumber\\
&&\lefteqn{\hspace{5.5cm}+Y^k_{ij}{Y^k_{ji}}^*\left(\frac{1}{12}+\frac{1}{2}\frac{{m_{\phi}}^2}{(m_{q_i}+m_{q_j})^2}\right)
\cdot B_{4}\cdot \eta_4(\mu)\Big].}\label{m12}
\end{eqnarray}
Here $\phi$ is the neutral meson $(K^0,~B_d^0, ~B_s^0,~D^0)$. For our numerical study we use the modified vacuum
saturation and factorization approximation results for the matrix
elements \cite{ciuchini,becirevic}
$$
\langle\phi|\bar{f}_i(1\pm\gamma_5)f_j\bar{f}_i(1\mp\gamma_5)f_j|\bar{\phi}\rangle={f_{\phi}}^2m_{\phi}
\left(\frac{1}{6}+\frac{{m_{\phi}}^2}{(m_{q_i}+m_{q_j})^2}\right)\cdot B_{4},
$$
\begin{equation}
\langle\phi|\bar{f}_i(1\pm\gamma_5)f_j\bar{f}_i(1\pm\gamma_5)f_j|\bar{\phi}\rangle
=-\frac{5}{6}{f_{\phi}}^2m_{\phi}\frac{{m_{\phi}}^2}{(m_{q_i}+m_{q_j})^2}\cdot B_{2}.
\end{equation}
$B_2$ and $B_4$ are equal to one in the vacuum saturation approximation, but are found
to be slightly different from one in lattice simulations.  We use  $(B_2,~B_4) = (0.66,~1.03)$ for
the $K^0$ system, $(0.82,~1.16)$ for the $B_d^0$  and $B_s^0$ systems, and $(0.82,~1.08)$ for the
$D^0$ system \cite{ciuchini}. In Eq. (\ref{m12})
$\eta_2(\mu)$, $\eta_4(\mu)$ are QCD correction factors of the Wilson coefficients $C_2$ and $C_4$ of the
effective $\Delta F = 2$ Hamiltonian in going from the SUSY scale $M_s$ to the hadronic scale $\mu$.
These factors are computed as follows.  The $\Delta F = 2$ effective Hamiltonian has the general form
\begin{equation}
\mathcal{H}_{\rm eff}^{\Delta F = 2}=\sum^5_{i=1}C_i~Q_i+\sum^3_{i=1}\tilde{C}_i~\tilde{Q},
\end{equation}
where
\begin{eqnarray}
Q_1={\bar{q_i}}^{\alpha}_L\gamma_{\mu}{q_j}^{\alpha}_L\bar{q_i}^{\beta}_L\gamma^{\nu}{q_j}^{\beta}_L, ~~~~
Q_2={\bar{q_i}}^{\alpha}_R {q_j}^{\alpha}_L \bar{q_i}^{\beta}_R {q_j}^{\beta}_R,~~~~
Q_3={\bar{q_i}}^{\alpha}_R {q_j}^{\beta}_L \bar{q_i}^{\beta}_R {q_j}^{\alpha}_L,\nonumber\\
Q_4={\bar{q_i}}^{\alpha}_R {q_j}^{\alpha}_L \bar{q_i}^{\beta}_L {q_j}^{\beta}_R,~~~~~~~~~~~
Q_5={\bar{q_i}}^{\alpha}_R {q_j}^{\beta}_L \bar{q_i}^{\beta}_L {q_j}^{\alpha}_R,\hspace{1.8 cm}
\end{eqnarray}
with $\tilde{Q}_{1,2,3}$ obtained from $Q_{1,2,3}$ by the interchange $L\leftrightarrow R$.

For computing $\eta_{2,4}$  we take the SUSY scale $M_s$ to be 1 TeV.  All the supersymmetric particles
and heavy Higgs bosons are integrated out at 1 TeV.  The
Wilson coefficients evolve from $M_s$ down to the hadron scale $\mu$ according to the equations
\begin{equation}
C_r(\mu)=\sum_i \sum_s (b_i^{(r,s)}+\eta c_i^{(r, s)})\eta^{a_i}C_s(M_s),
\end{equation}
Here $\eta$ is defined as $\eta=\alpha_s(M_s)/\alpha_s(m_t)$. The magic numbers $a_i$, $b_i^{(r, s)}$ and $c_i^{(r, s)}$ can be found in Ref. \cite{ciuchini} for the $K$ system, in Ref.  \cite{becirevic} for the $B_{d, s}$ system and
in Ref. \cite{utfit} for the $D$ system.
With $M_s=1 {\rm ~TeV}$  and $\alpha_s(m_Z) = 0.118$, and $m_t(m_t)=163.6 {\rm ~GeV}$ we find
$\eta=\alpha_s(1 {\rm ~TeV})/\alpha_s(m_t)=0.0882/0.108=0.8167$.

At the SUSY scale, the neutral Higgs bosons
in our model generate only operators $Q_2$ and $Q_4$.   Consequently, at the hadron scale, for the $K^0$ system,
we find
\begin{eqnarray}
C_2(\mu)=C_2(M_s)\cdot  (2.54), ~~~~~C_4(\mu)=C_4(M_s)\cdot  (4.81),\nonumber\\
C_3(\mu)=C_2(M_s) \cdot (-1.8\times10^{-3}),~~~~~ C_5(\mu)=C_4(M_s)\cdot (0.186),
\end{eqnarray}
leading to $\eta_2(\mu) = 2.54,~\eta_4(\mu) = 4.81$.  Although operator mixings induce
non-zero $C_3$ and $C_5$ at the hadronic scale, their coefficients are found to be rather
small.

For the $B^0_{d, s}$ system, following the same procedure, we find
\begin{eqnarray}
C_2(\mu)=C_2(M_s)\cdot (2.00), ~~~~~C_4(\mu)=C_4(M_s)\cdot  (3.12),\nonumber\\
C_3(\mu)=C_2(M_s) \cdot (-2.44\times10^{-2}),~~~~~ C_5(\mu)=C_4(M_s)\cdot (0.0874).
\end{eqnarray}

And for the $D^0$ system we have
\begin{eqnarray}
C_2(\mu)=C_2(M_s)\cdot (2.31), ~~~~~C_4(\mu)=C_4(M_s)\cdot  (3.99),\nonumber\\
C_3(\mu)=C_2(M_s) \cdot (-1.30\times10^{-2}),~~~~~ C_5(\mu)=C_4(M_s)\cdot (0.144).
\end{eqnarray}
In all cases we see that the induced operators $C_3$ and $C_5$ are negligible.

\vspace*{0.1in}
\noindent {\bf $K^0-\overline{K^0}$ mixing constraint}:
\vspace*{0.1in}

In the $K^0$ system,  tree--level neutral Higgs boson exchange contributes to $K_L-K_S$ mass difference,
as well as to the indirect CP violation parameter, modifying the successful SM predictions.  The mass difference is computed from
$\Delta m_{K}=2{\rm Re} M^{K}_{12}$, while the CP violation parameter is
$|\epsilon_K|\simeq\frac{{\rm Im} M^{K}_{12}}{\sqrt{2}\Delta m_{K}}$.  We
seek consistency with the precisely measured experimental values $\Delta m_K/m_K\simeq (7.1\pm 0.014)\times10^{-15}$
and  $|\epsilon_K|\simeq 2.3\times10^{-3}$.  In our calculation, we choose $m_K=498 {~\rm MeV}$ and $f_K=160{~\rm MeV}$. For the two numerical fits we find the new contributions to be
\begin{eqnarray}
{\rm Case}~ {\bf (1)}: ~~(\Delta m_K/m_K)^{\rm new}=7.361\times 10^{-15}, ~~\epsilon_K^{\rm new}=2.00\times 10^{-4},
\nonumber \\
{\rm Case}~ {\bf (2)}: ~~(\Delta m_K/m_K)^{\rm new}=5.721\times 10^{-15}, ~~\epsilon_K^{\rm new}=2.28\times 10^{-5}.
\end{eqnarray}
The contributions from $H_1^0-H_3^0$ sector and $H_2^0$ sector  to Re$(M_{12}^K)$ are respectively
$(3.033\times 10^{-15}$, $-1.200\times 10^{-15}$) GeV for case (1) and
($2.512\times 10^{-15}$, $-1.088\times 10^{-15}$) GeV for case (2).
We see that the new contributions to the mass difference is significant, but consistent with data.
New contributions to CP violation is suppressed, which is a generic feature of Higgs exchange in this
class of models.  We elaborate on this issue later in this section.

\vspace*{0.1in}
\noindent {\bf $B_d^0-\overline{B_d^0}$ mixing constraint}:
\vspace*{0.1in}

For the $B_d^0-\bar{B_d^0}$ system We use as input $m_{B_d}=5.281{~\rm GeV}, ~f_{B_d}=240{~\rm MeV}$ and
seek consistency with the experimental value $\Delta m_{B_d}=3.12\times 10^{-13}{~\rm GeV}$. We find
for the Higgs induced contribution
\begin{eqnarray}
{\rm Case}~ {\bf (1)}:~~(\Delta m_{B_d})^{\rm new}=2.997\times 10^{-13}{~\rm GeV}, \nonumber \\
{\rm Case}~ {\bf (2)}:~~(\Delta m_{B_d})^{\rm new}=2.728\times 10^{-13}{~\rm GeV}.
\end{eqnarray}
The contributions from $H_1^0-H_3^0$ sector and $H_2^0$ sector to $M_{12}^{b_d}$ are  $(2.298\times 10^{-14}$, $1.269\times 10^{-13}$) GeV for case (1) and
$(2.137\times 10^{-14}$, $1.150\times 10^{-13}$) GeV.
Again, we see consistency with experimental values.  CP violation parameter is found to be extremely
tiny, $\sim 10^{-5}$, from Higgs boson exchange.

\vspace*{0.1in}
\noindent {\bf $B_s^0-\overline{B_s^0}$ mixing constraint}:
\vspace*{0.1in}

For the $B_s^0-\overline{B_s^0}$ system, we use $m_{B_s}=5.37{~\rm GeV}, ~f_{B_s}=295{~\rm MeV}$
and compare the new contributions with $\Delta m_{B_s}=1.067\times 10^{-11}{~\rm GeV}$.
\begin{eqnarray}
{\rm Case}~ {\bf (1)}:~~(\Delta m_{B_s})^{\rm new}=1.688\times 10^{-12}{~\rm GeV}, \nonumber \\
{\rm Case}~ {\bf (2)}:~~(\Delta m_{B_s})^{\rm new}=1.396\times 10^{-12}{~\rm GeV},
\end{eqnarray}
The $H_1^0-H_3^0$ sector and the $H_2^0$ sector contribute to $M_{12}^{B_s^0}$ given by
$(8.532\times 10^{-13}$, $-9.460\times 10^{-15}$) GeV for case (1) and
$(7.067\times 10^{-13}$, $-3.835\times 10^{-15}$) GeV for case (2).  These new contributions are
within experimentally allowed range.  Higgs mediated CP violation is again found to be highly suppressed.

\vspace*{0.1in}
\noindent {\bf $D^0-\overline{D^0}$ mixing constraint}:
\vspace*{0.1in}

For the $D^0-\overline{D^0}$ mixing we use $m_D=1.864 {\rm ~GeV}$, $f_D=200{\rm ~MeV}$ and compare the new contribution with
$\Delta m_{D}=1.27\times 10^{-12}{~\rm GeV}$.
\begin{eqnarray}
{\rm Case}~ {\bf (1)}:~~(\Delta m_D)^{\rm new}=8.620\times 10^{-13}~ {\rm GeV}, \nonumber \\
{\rm Case}~ {\bf (2)}:~~(\Delta m_D)^{\rm new} =2.645\times 10^{-13}~ {\rm GeV},
\end{eqnarray}
The $H_1^0-H_3^0$ sector contribution has different sign from that of the $H_2^0$ sector. We
find for $M_{12}^{D^0}$ these contributions to be ($4.402\times 10^{-15}$, $-4.354\times 10^{-13}$) GeV for case (1) and $(2.568\times 10^{-15}$, $-1.348\times 10^{-13})$ GeV for case (2).  Again these
limits are within experimental range.

We have found that new sources of
CP violation through tree--level Higgs  is very small in meson--antimeson mixings with typical values
 Im$(M_{12}) \sim
10^{-4}$ Re$(M_{12}$). This can  be understood heuristically
as follows. There are two types of contributions to the meson mixing as given in Eq. (\ref{m12}).  The first
term, proportional to $B_2$ respects a global $U(1)$ symmetry (strangeness in the $K^0$ system), which is
only broken by the mass--splittings in the neutral Higgs boson spectrum between a pair of particles.
However, this splitting is very small, of order $m_Z^2$ in the squared mass, see Eqs. (\ref{spectrum1})
.  The couplings of the nearly degenerate Higgs in each pair differ by a factor $i$,
owing to the $U(1)$ symmetry, and the two contributions cancel, in the limit of exact degeneracy.
For both the real and
imaginary parts of $M_{12}$ the contributions from the first term is suppressed by a factor $m_Z^2/(4 M_k^2)$.
Such a suppression is absent in the second term of Eq. (\ref{m12}), since the operator $Q_4$ explicitly
breaks the $U(1)$ symmetry.  Thus, although the first term has CP violation, in relation to the CP--conserving
second term, it is suppressed by a factor $m_Z^2/(4 M_k^2) \sim 10^{-4}$.  Now, the second term, while it has
no suppression factor, it is purely real.  This can be seen from the following observation.
In the mass basis of fermions in the original basis we have the relation (owing to the vanishing of
off--diagonal mass terms in the mass eigenbasis)
\begin{equation}
(Y_{d3})_{ij}\left<H^d_3\right>=-(Y_{d1})_{ij}\left<H^d_1\right>
\end{equation}
for $i\neq j$. The couplings of mass eigenstates of Higgs boosn to down--type quarks are simply linear combinations of $H^d_1$ and $H^d_3$.  Since we assume CP to be spontaneously broken, all components of $(Y^k)_{ij}$ with $i\neq j$ have the same phase. As a result the second term of Eq. \ref{m12} becomes real.  The constraint imposed by SUSY,
that $H_u^*$ fields do not couple to down--type quarks, and the fact that only two of the down--type Higgs bosons
acquire VEVs is very crucial for this result.

\subsection{Neutron electric diploe moment from Higgs exchange}

Higgs boson exchange can generate non-zero electric dipole moments for the fermions. These diagrams
are however suppressed by the light fermion Yukawa couplings.  For the
$d$ quark EDM arising from neutral Higgs boson exchange at the one--loop level we find \cite{he}
\begin{equation}
d_d=\frac{Q_d e}{16\pi^2}{\rm Im}(Y^k_{dq}Y^k_{qd})\frac{m_q}{M_k^2}\left[\frac{3}{2}-\ln\left(\frac{M_k^2}{m_q^2}\right)\right]\xi_d,
\end{equation}
where $\xi_d=(\alpha_s(M_k)/\alpha_s(\mu))^{16/23}\approx0.12$, and $q$ is summed over $d,~s$ and $b$. The neutron EDM is determined using the quark model via
 \begin{equation}
D_n=4d_d/3-d_u/3.
\end{equation}
We find
\begin{eqnarray}
{\rm Case ~(1)}:~~~D_n=1.809\times 10^{-31}~ {~\rm e-cm}, \nonumber \\
{\rm Case ~(2)}:~~~D_n=6.091\times 10^{-31}~ {~\rm e-cm},
\end{eqnarray}
which are well within experimental limits.  The EDM of the electron is similarly found to be extremely
small from the Higgs boson exchange diagrams.

\subsection{$\mu\rightarrow3e$ and $\tau \rightarrow 3 \mu$ decays}

Tree--level Higgs boson exchange can lead to flavor violating leptonic decays such as $\tau \rightarrow
3 \mu$ and $\mu \rightarrow 3 e$.
The effective weak interaction mediating such decays can be parametrized as
\begin{equation}
G_{\rm eff}= \left|\sum_k (Y_e)^k_{11}(Y_e)^k_{12}\frac{1}{M_k^2}\right|
\end{equation}
The effective couplings are found for $\mu \rightarrow 3 e$ for the two cases to be
\begin{eqnarray}
{\rm Case ~(1)}:~~~G_{\rm eff}=4.432\times 10^{-13}~G_F, \nonumber \\
{\rm Case ~(2)}:~~~G_{\rm eff}=4.191\times 10^{-13}~G_F~.
\end{eqnarray}
And the couplings for $\tau\rightarrow 3\mu$ decay are
\begin{eqnarray}
{\rm Case ~(1)}:~~~G_{\rm eff}=45.721\cdot 10^{-8}~G_F, \nonumber \\
{\rm Case ~(2)}:~~~G_{\rm eff}=6.977\cdot 10^{-8}~G_F~.
\end{eqnarray}
Such small effective couplings will lead to negligible contributions to
the decay branching ratios.  For example, the branching ratio for
$\tau \rightarrow 3 \mu$ is of order $10^{-15}$, well below the experimental
sensitivity.  We conclude that Higgs mediated FCNC in the lepton sector
are all safe.

\section{FCNC mediated by SUSY particles}

In this section we turn attention to the flavor changing processes mediated by
the supersymmetric particles.  The main motivation for the non--Abelian $Q_6$ model
was to bring such processes under control by a symmetry reason.  Here we analyze
meson--antimeson mixings, flavor violating leptonic decays, and the EDM of the
neutron and the electron.  We present our proposal to suppress SUSY contributions
to the EDM by making the Higgsinos of the model light, with masses of order $100$ GeV.

Owing to the $Q_6$ symmetry, the first two family squarks (and similarly sleptons) are
degenerate in mass, while the third family, which is a $Q_6$ singlet has a different mass.
In the fermion sector $Q_6$ symmetry is broken, which means that there will be SUSY loop
induced flavor violation in the model.  Constraints on such flavor violation has been
listed in Ref. \cite{ciuchini,becirevic,utfit} assuming all three families of squarks
are degenerate.  While these results are applicable for the $K^0$ and $D^0$ system in
our model, they do not work well for the $B_{d,s}^0$ system.  This is because the masses
of the $\tilde{b}$ and $\tilde{d,s}$ masses are not the same.

\subsection{Generalized constraints for $B_d$ system}

We have generalized the results of Ref. \cite{becirevic} by allowing for $\tilde{b}$
mass to be different from the masses of $\tilde{d,s}$.  We define new parameters
\begin{equation}
y^d_{A,B} = {(\tilde{m_b}^2)_{A,B} \over \tilde{m_d}^2_{A,B}}
\end{equation}
for $A,B = L,R$.  We expect these $y$ parameters to be of order one, but not very close
to one.  Taking account of $y\neq 1$ we have generalized the constraints on the squark
mixing parameters from $B_d^0$ system as follows.

The effective $\Delta F=2$ Hamiltonian for $B_{d,s}$ system can be written as
\begin{eqnarray}
&&\lefteqn{\mathcal{H}_{\rm eff}=\sum^5_{i=1}C_i~Q_i+\sum^3_{i=1}\tilde{C}_i~\tilde{Q}}\hspace{17cm}\nonumber\\
&&\lefteqn{\hspace{1cm}=-\frac{\alpha_s}{216 m^2_{\tilde{d}}}
\{(\delta^d_{13})^2_{LL}(24Q_1 x f_6(x, y)+66Q_1\tilde{f}_6(x, y))+
(\delta^d_{13})^2_{RR}(24\tilde{Q}_1 x f_6(x, y)+66\tilde{Q}_1\tilde{f}_6(x, y))}\nonumber\\
&&\lefteqn{\hspace{1cm}+(\delta^d_{13})_{LL}(\delta^d_{13})_{RR}(504 Q_4 x f_6(x, y)
-72Q_4\tilde{f}_6(x, y)+24 Q_5 xf_6(x, y)+120 Q_5 \tilde{f}_6(x, y))}\nonumber\\
&&\lefteqn{\hspace{1cm}+(\delta^d_{13})^2_{RL}(204 Q_2 x f_6(x, y)-36Q_3 x f_6(x, y))
+(\delta^d_{13})^2_{LR}(204 \tilde{Q}_2 x f_6(x, y)-36\tilde{Q}_3 x f_6(x, y))}\nonumber\\
&&\lefteqn{\hspace{1cm}+(\delta^d_{13})_{LR}(\delta^d_{13})^2_{RL}(-132 Q_4 \tilde{f}(x, y)-180 Q_5\tilde{f}_6(x, y))\},}
\end{eqnarray}

The functions $f_6(x, y)$ and $ \tilde{f}_6(x, y)$ are
\begin{eqnarray}
&&\lefteqn{ \hspace{-8 cm}f_6(x, y)=\frac{1}{(x-1)^3(y-1)^3(z-1)^3}\Big[-\ln {x}(x+y+xy-3x^3)(y-1)^3 }\nonumber\\
&&\lefteqn{\hspace{-5 cm} +\ln{y}(x+y+xy-3y^2)(x-1)^3 } \nonumber\\
&&\lefteqn{\hspace{-5 cm} +2(x-1)(y-1)(-x+y+x^2-y^2-x^3+y^3+2x^2y-2xy^2) \Big]}\nonumber\\
&&\lefteqn{ \hspace{-8 cm}\tilde{f}_6(x, y)=\frac{1}{(x-1)^3(y-1)^3(z-1)^3}\Big[2\ln {x}\cdot x(x^2-y)(y-1)^3 }\nonumber\\
&&\lefteqn{\hspace{-5 cm} +2\ln{y}\cdot y(x-y^2)(x-1)^3 } \nonumber\\
&&\lefteqn{\hspace{-5 cm} +(x-1)(y-1)(x^2-y^2+x^3-y^3-7x^2y+7xy^2+x^3y-xy^3)\Big].}
\end{eqnarray}

Generalizing the results of Ref. \cite{becirevic} we obtain the squark mixing coefficients $(\delta^d_{13})_{AB}$ with $A,~B=(L,~R)$ as  shown in Table 1.  Here we have used the same input as in Ref. \cite{becirevic}, so that for
$y=1$ our results coincide.  We have used the next-to-leading order lattice calculation results for the
matrix elements.  For some of the mixing parameters we made a simplifying assumption that $y_L^d$ and
$y_R^d$ are equal.

\begin{center}
\begin{table}[h]
\hspace{1.2cm}
\begin{tabular}{|| c|| c |c| c | c |c| c ||}
\hline\hline
$y\diagdown x$&0.25&1.0&4.0&0.25&1.0&4.0\\
\hline
&\multicolumn{3}{c|}{$|{\rm Re}(\delta^d_{13})_{\rm LL}|$}  & \multicolumn{3}{|c||}{$|{\rm Im}(\delta^d_{13})_{\rm LL}|$} \\
\hline
0.25&$3.4\times 10^{-2}$&$1.6\times 10^{-1}$&$2.5\times 10^{-1}$&$7.2\times 10^{-2}$&$3.4\times 10^{-1}$&$1.2\times 10^{-1}$\\
\hline
1.0&$6.2\times 10^{-2}$&$1.4\times 10^{-1}$&$7.0\times 10^{-1}$&$1.3\times 10^{-1}$&$3.0\times 10^{-1}$&$3.4\times 10^{-1}$\\
\hline
4.0&$1.6\times 10^{-1}$&$2.7\times 10^{-1}$&|&$3.3\times 10^{-1}$&$5.8\times 10^{-1}$&|\\
\hline
&\multicolumn{3}{c}{$|{\rm Re}(\delta^d_{13})_{\rm RR}|=|{\rm Re}(\delta^d_{13})_{\rm LL}|$}\
& \multicolumn{3}{|c||}{$|{\rm Im}(\delta^d_{13})_{\rm RR}|=|{\rm Im}(\delta^d_{13})_{\rm LL}|$}\\
\hline
0.25&$1.4\times 10^{-3}$&$2.4\times 10^{-2}$&$1.0\times 10^{-2}$&$4.4\times 10^{-3}$&$1.0\times 10^{-2}$&$4.3\times 10^{-3}$\\
\hline
1.0&$1.9\times 10^{-2}$&$2.1\times 10^{-2}$&$2.8\times 10^{-2}$&$8.0\times 10^{-3}$&$9.0\times 10^{-1}$&$1.2\times 10^{-2}$\\
\hline
4.0&$4.8\times 10^{-2}$&$4.0\times 10^{-2}$&$1\times 10^{-1}$&$2\times 10^{-2}$&$1.7\times 10^{-2}$&$4.6\times 10^{-2}$\\
\hline
&\multicolumn{3}{c}{$|{\rm Re}(\delta^d_{13})_{\rm LR}|$} & \multicolumn{3}{|c||}{$|{\rm Im}(\delta^d_{13})_{\rm LR}|$}\\
\hline
0.25&$1.7\times 10^{-2}$&$3.7\times 10^{-2}$&$1.6\times 10^{-2}$&$3.6\times 10^{-2}$&$8.4\times 10^{-2}$&$3.6\times 10^{-2}$\\
\hline
1.0&$3.0\times 10^{-2}$&$3.3\times 10^{-2}$&$4.5\times 10^{-2}$&$6.6\times 10^{-2}$&$7.4\times 10^{-2}$&$1.0\times 10^{-1}$\\
\hline
4.0&$7.5\times 10^{-2}$&$6.4\times 10^{-2}$&$1.7\times 10^{-1}$&$1.7\times 10^{-1}$&$1.4\times 10^{-1}$&$3.9\times 10^{-1}$\\
\hline
&\multicolumn{3}{c}{$|{\rm Re}(\delta^d_{13})_{\rm LR}|=|{\rm Re}(\delta^d_{13})_{\rm RL}|$}&
\multicolumn{3}{|c||}{$|{\rm Im}(\delta^d_{13})_{\rm LR}|=|{\rm Im}(\delta^d_{13})_{\rm RL}|$}\\
\hline
0.25&$1.4\times 10^{-2}$&$5.9\times 10^{-2}$&|&$2.3\times 10^{-2}$&$4.4\times 10^{-1}$&|\\
\hline
1.0&$2.6\times 10^{-2}$&$5.2\times 10^{-2}$&|&$9.0\times 10^{-3}$&$2.3\times 10^{-2}$&|\\
\hline
4.0&$6.5\times 10^{-2}$&$1.0\times 10^{-1}$&|&$2.3\times 10^{-2}$&$4.4\times 10^{-2}$&|\\
\hline\hline
\end{tabular}

\caption{Maximum allowed values for $|{\rm Re}(\delta^d_{13})_{\rm AB}|$ and $|{\rm Im}(\delta^d_{13})_{\rm AB}|$, with $A, B=(L, R)$.
A new parameter $y$ is introduced, with $y=m^2_{\tilde{b}}/m^2_{\tilde{d}}$. The definition
of other parameters and their values
follow Ref. \cite {becirevic }.}

\end{table}
\end{center}

\subsection {SUSY flavor change in $Q_6$ model}
 In the $Q_6$ model the  mass matrices of squarks in the flavor basis can be written as
\begin{eqnarray}
(m_{\tilde{q}})^2_{AA}=m^2_{\tilde{q}A}\left(\begin{array}{ccc}1&0&0\\0&1&0\\0&0&y\end{array}\right),
\end{eqnarray}
$q$ can be $u$ or $d$, $A$ can be $L$ or $R$. Making the same unitary transformation on the squark fields
as the ones on the  quarks which diagonalize the quark mass matrices, we find the mass
matrices of squarks in the SUSY basis (where the gluino coupling matrix is identity in the flavor space) to be:
\begin{eqnarray}
(\tilde{m}_{\tilde{d}})^2_{LL}=O_d^T P_d^{\ast}(m_{\tilde{d}})^2_{LL}P_d O_d
=m^2_{\tilde{d}L}\left[I+(y^d_l-1)O_d^T P_d^{\ast}\left(\begin{array}{ccc}0&0&0\\0&0&0\\0&0&1\end{array}\right)P_d O_d\right]\nonumber\\
=m^2_{\tilde{d}L}\left[I+(y^d_L-1)\left(\begin{array}{ccc}
5.43\cdot 10^{-5}&1.31\cdot 10^{-4}&-0.0074\\1.31\cdot 10^{-4}&3.17\cdot 10^{-4}&-0.0178\\-0.0074&-0.0178&0.9996\end{array}\right)\right],
\end{eqnarray}
Note that this matrix is real, a consequence of
the phase factorization of the fermion mass matrix.  Similarly,
\begin{eqnarray}
(\tilde{m}_{\tilde{d}})^2_{RR}=m^2_{\tilde{d}R}\left[I+(y^d_R-1)\left(\begin{array}{ccc}
0.0367&-0.1339&0.1318\\-0.1339&0.4891&-0.4816\\0.1319&-0.4816&0.4742\end{array}\right)\right],
\end{eqnarray}

\begin{eqnarray}
(\tilde{m}_{\tilde{u}})^2_{LL}=m^2_{\tilde{u}L}\left[I+(y^u_L-1)\left(\begin{array}
{ccc}3.85\cdot 10^{-6}&7.74\cdot 10^{-5}&-0.0020\\7.74\cdot 10^{-5}&0.0016&-0.0394\\-0.0020&-0.0394&0.9984
\end{array}\right)\right],
\end{eqnarray}

\begin{eqnarray}
(\tilde{m}_{\tilde{u}})^2_{RR}=m^2_{\tilde{u}R}\left[I+(y^u_R-1)\left(\begin{array}{ccc}
1.46\cdot 10^{-5}&2.94\cdot 10^{-4}&0.0038\\2.94\cdot 10^{-4}&0.0059&-0.0768\\0.0038&-0.0768&0.9941\end{array}\right)\right].
\end{eqnarray}

$K^0-\bar{K^0}$ mixing via squark--gluino loops have several contributions.  The most stringent
limit arises from the $(LL)-(RR)$ mixing, which requires \cite{ciuchini}
\begin{equation}
{|(y^d-1)| \over (0.51 + 0.49 ~y^d)^{1/4}} ~<~ 0.23 \left({\tilde{m} \over 500~{\rm  GeV}}\right)~.
\end{equation}
Here we have assumed $y^d_L = y^d_R = y^d$, and took the gluino mass to be equal to the
first two family squrak mass.  For first two family squark mass of 500 GeV, this translates to
the limit $0.77 \leq y^d \leq 1.24$.  For 1 TeV squarks, this limit is relaxed to $0.58 \leq y^d \leq 1.48$.
We see that for $y^d$ order one, the most stringent limit on squark mediated FCNC is satisfied.

The $Q_6$ model also generates significant $(RR)(RR)$ contributions to the $K^0-\overline{K^0}$ mixing.  We find
\begin{equation}
0.68 \leq y^d \leq 1.37
\end{equation}
for squark and gluino  mass of 500 GeV.  This constraint is also easily satisfied in the model.

In the $B_d^0$ system, the analogous constraints are (from the $(LL)(RR)$ operator)
\begin{equation}
{|(y^d-1)| \over (0.53 + 0.47 ~y^d)^{1/4}} ~<~ 0.69 \left({\tilde{m} \over 500~{\rm  GeV}}\right)~.
\end{equation}
This limit leads to $0.48 \leq y^d \leq 1.85$ for squark-gluino mass of 500 GeV.  The $(RR)(RR)$ squark
mixing gives no constraint from the $B_d$ system.  Similarly, there are no constraints arising from
the $D^0$ system, nor from other type of operators in the model.

In the leptonic sector, we find the $(LL)$ slepton mixing (which is the same for the (RR) slepton mixing)
to be
\begin{eqnarray}
(\tilde{m}_{\tilde{e}})^2_{LL}=m^2_{\tilde{e}L}\left[I+(y^e_L-1)\left(\begin{array}{ccc}
2.93\cdot 10^{-4}&-4.02 \cdot 10^{-3}&-0.0167\\-4.02 \cdot 10^{-3}&0.0550&0.2280\\-0.0167&0.2280&0.9447\end{array}\right)\right]~.
\end{eqnarray}

There are stringent constraints on the mixing parameter $((\delta^e)_{LL})_{12}$ from the decay
$\mu \rightarrow e \gamma$ \cite{vempati}.  On the face of it, the mixing presented above would appear
to be in mild conflict with data by a factor of few.  However, since such a constraint is very week for the
$((\delta^e)_{RR})_{12}$ mixing, we point out that the flexibility in the lepton sector mass matrix
can be used to make the $(LL)$ contribution small in exchange for larger $(RR)$ contributions.
That is, assume $B \ll B'$ in Eq. (\ref{massmatrix}).

\subsection{Left--Right squark mixing and a solution to the EDM problem}

So far we have ignored SUSY flavor violation arising from the left--right squark mixings.  It turns out
that these operators do not give significant contributions to meson--antimeson mixings, since
such mixings have fermion chirality suppression.  However, these mixings can generate new contributions
to the neutron (and electron) electric dipole moments.  Here we analyze constraints from the EDM and
suggest a simple solution to the SUSY EDM problem.

First, as shown in Ref. \cite{babu}, the trilinear $A$--term induced phases align with the phases of
the fermion mass matrices, even without assuming proportionality of the $A$--terms with the respective
Yukawa couplings.  This feature arises due to the phase factorization of the fermion mass matrix.
Left--right squark mixings also receive contributions from the superpotential $\mu$-terms. We derive
the mass matrix for the down squark sector to be:
\begin{eqnarray}
(m_{\tilde{d}})^2_{LR}={F^d_1}^{\ast} \left(\begin{array}{ccc}0&\delta_d&0\\-\delta_d&0&0\\0&0&\alpha_d \end{array}\right)
+\sqrt{2} {F^d_2}^{\ast} \left(\begin{array}{ccc}0&0&0\\0&0&\beta_d\\0&\beta^{'}_d&0 \end{array}\right),
\end{eqnarray}
with
\begin{eqnarray}
F^d_1=\mu_3 v_{u3}+\mu_{13}v_{u1},~~~~~~~~~F^d_2=\mu_{31}v_{u3}+\frac{\mu_1+\mu_{12}}{2}v_{u1}.
\end{eqnarray}

After the unitary transformations to the left and the right squarks, corresponding to case (1), we have
the $(LR)$ mixing matrix in the flavor basis as
\begin{eqnarray}
&&(\tilde{m}_{\tilde{d}})^2_{LR}=O_d^T P_d(m_{\tilde{d}})^2_{LR}P_{d^c} O_{d^c}
={F^d_1}^{\ast}\left(\begin{array}{ccc}
-1.75 \cdot 10^{-4}&4.14 \cdot 10^{-4}&3.09 \cdot 10^{-5}\\
-4.46\cdot 10^{-4}&3.84\cdot 10^{-4} &-5.45\cdot 10^{-4} \\
0.0078&-0.0286&0.0282 \end{array}\right)\nonumber\\
&&\hspace{3cm}+ \sqrt{2} {F^d_2}^{\ast} e^{i~\Delta \theta_d}
\left(\begin{array}{ccc}
4.53\cdot 10^{-5}&-1.66\cdot 10^{-4}&-1.24\cdot 10^{-5}\\
1.79\cdot 10^{-4}&-6.52\cdot 10^{-4}&2.18\cdot 10^{-4}\\
-0.0031&0.0115&0.0125 \end{array}\right),\hspace{1cm}
\end{eqnarray}

\begin{eqnarray}
&&(\tilde{m}_{\tilde{u}})^2_{LR}=O_u^T P_u(m_{\tilde{u}})^2_{LR}P_{u^c} O_{u^c}
={F^u_1}^{\ast} \left(\begin{array}{ccc}
-1.62\cdot 10^{-5}&2.18\cdot 10^{-4}&-0.0014\\
-2.18\cdot 10^{-4}&0.0022&-0.0283\\
0.0027&-0.0554&0.7168 \end{array}\right)\nonumber\\
&&\hspace{3cm}+ \sqrt{2} {F^u_2}^{\ast} e^{i~\Delta \theta_u}
\left(\begin{array}{ccc}
3.24\cdot 10^{-5}&-6.53\cdot 10^{-4}&0.0042\\
6.53\cdot 10^{-4}&-0.0131&0.0848\\
-0.0082&0.1661&0.0162 \end{array}\right).\hspace{1cm}
\end{eqnarray}
with
\begin{eqnarray} {F^u_1}=\mu_3 v_{d3}+\mu_{31}v_{d1},~~~~~~~~~{F^u_2}=\mu_{13}v_{d3}+\frac{\mu_1+\mu_{12}}{2}v_{d1}.
\end{eqnarray}

\begin{eqnarray}
&&(\tilde{m}_{\tilde{e}})^2_{LR}=O_e^T P_e(m_{\tilde{e}})^2_{LR}P_{e^c} O_{e^c}
={F^d_1}^{\ast}\left(\begin{array}{ccc}
-1.10 \cdot 10^{-5}&2.85 \cdot 10^{-4}&6.72 \cdot 10^{-4}\\
-2.75\cdot 10^{-4}&0.0039 &0.0257 \\
-4.89\cdot 10^{-5}&0.0047&0.0313 \end{array}\right)\nonumber\\
&&\hspace{3cm}+ \sqrt{2} {F^d_2}^{\ast} e^{i~\Delta \theta_d}
\left(\begin{array}{ccc}
2.93\cdot 10^{-6}&-1.14\cdot 10^{-4}&-2.69\cdot 10^{-4}\\
1.10\cdot 10^{-4}&-0.0043&-0.0103\\
1.96\cdot 10^{-5}&-0.0019&0.0093 \end{array}\right),\hspace{1cm}
\end{eqnarray}

Corresponding to case (2) these matrices are:
\begin{eqnarray}
&&(\tilde{m}_{\tilde{d}})^2_{LR}=O_d^T P_d(m_{\tilde{d}})^2_{LR}P_{d^c} O_{d^c}
={F^d_1}^{\ast}\left(\begin{array}{ccc}
-2.25\cdot 10^{-4}&5.32 \cdot 10^{-4}&3.97 \cdot 10^{-5}\\
-5.73\cdot 10^{-4}&4.93\cdot 10^{-4} &-7.00\cdot 10^{-4} \\
0.0100&-0.0368&0.0362
 \end{array}\right)\nonumber\\
&&\hspace{3cm}+ \sqrt{2} {F^d_2}^{\ast} e^{i~\Delta \theta_d}
\left(\begin{array}{ccc}
5.20\cdot 10^{-5}&-1.90\cdot 10^{-4}&-1.42\cdot 10^{-5}\\
2.05\cdot 10^{-4}&-7.48\cdot 10^{-4}&2.50\cdot 10^{-4}\\
-0.0036&0.0131&0.0144
 \end{array}\right),\hspace{1cm}
\end{eqnarray}

\begin{eqnarray}
&&(\tilde{m}_{\tilde{u}})^2_{LR}=O_u^T P_u(m_{\tilde{u}})^2_{LR}P_{u^c} O_{u^c}
={F^u_1}^{\ast} \left(\begin{array}{ccc}
-2.11\cdot 10^{-5}&2.83\cdot 10^{-4}&-0.0018\\
-2.83\cdot 10^{-4}&0.0028&-0.0368\\
0.0036&-0.0720&0.9319
\end{array}\right)\nonumber\\
&&\hspace{3cm}+ \sqrt{2} {F^u_2}^{\ast} e^{i~\Delta \theta_u}
\left(\begin{array}{ccc}
1.41\cdot 10^{-5}&-2.83\cdot 10^{-4}&0.0018\\
2.83\cdot 10^{-4}&-0.0057&0.368\\
-0.0036&0.0720&0.0070
\end{array}\right).\hspace{1cm}
\end{eqnarray}

\begin{eqnarray}
&&(\tilde{m}_{\tilde{e}})^2_{LR}=O_e^T P_e(m_{\tilde{e}})^2_{LR}P_{e^c} O_{e^c}
={F^d_1}^{\ast}\left(\begin{array}{ccc}
-1.41 \cdot 10^{-5}&3.66 \cdot 10^{-4}&8.62 \cdot 10^{-4}\\
-3.53\cdot 10^{-4}&0.0050 &0.0330 \\
-6.28\cdot 10^{-5}&0.0061&0.0402
\end{array}\right)\nonumber\\
&&\hspace{3cm}+ \sqrt{2} {F^d_2}^{\ast} e^{i~\Delta \theta_d}
\left(\begin{array}{ccc}
3.36\cdot 10^{-6}&-1.31\cdot 10^{-4}&-3.08\cdot 10^{-4}\\
1.26\cdot 10^{-4}&-0.0049&-0.0118\\
2.24\cdot 10^{-5}&-0.0022&0.0106
\end{array}\right),\hspace{1cm}
\end{eqnarray}
Note that these matrices are in general complex, since $F_i^{u,d}$ are complex because of the
spontaneously induced phases of the VEVs.  This means that these matrices will contribute
to neutron and electron EDM.  Since these complex coefficients are proportional to $\mu v/\tilde{m}^2$,
we find a simple solution to the SUSY EDM problem:  Let the $\mu$ terms be of order 100 GeV,
in which case one finds a suppression factor of $10^{-2}$ for the effective phase that enters
the EDM expression.  With this suppression factor, from the $(1,1)$ elements of these $(LR)$ mixing
matrices, we see that neutron and electron EDM constraints can be satisfied, even with the
spontaneously induced phases in the VEVs being of order one.

The proposed solution to the SUSY EDM problem has direct experimental consequences for LHC.
We predict that the Higgsinos should be light, and three such pairs of doublet Higgsinos
should be observable at the LHC.  Their scalar partners, however, are inaccessible, since
their masses lie in the few TeV range.

\section{Conclusion}

In conclusion, we have presented a detailed analysis of the Higgs potential involving three pairs of
Higgs doublets in a $Q_6$ model of flavor.  This class of models are motivated on two grounds: They
lead to reduced number of parameters in the fermionic sector, and they can be helpful in alleviating
the flavor changing problems of generic SUSY models.

We have shown that tree--level Higgs boson induced FCNC are within experimental limits, even for
the most stringent $K^0-\overline{K^0}$ mixing amplitude.  The Higgs boson masses must lie in the
TeV range.  New sources of CP violation in meson mixing are highly suppressed.  We have also shown
the consistency of the model with SUSY flavor violation.  A simple solution to the SUSY EDM problem
is suggested, which requires light Higgsinos.

\subsection* {Acknowledgments}

\hs{-0.1cm} We have benefitted from discussions with Xiao-Gand He, Jisuke Kubo and Zurab Tavartkiladze.
Analysis similar to ours has recently been done by K.~Kawashima, J.~Kubo and A.~Lenz, arXiv:0907.2302 [hep-ph]. This work is supported in part by US Department of Energy, Grant Numbers
DE-FG02-04ER41306 and DE-FG02-ER46140.


\begin{thebibliography}{99}


\bibitem{susyfcnc}
  J.~F.~Donoghue, H.~P.~Nilles and D.~Wyler,
  Phys.\ Lett.\  B {\bf 128}, 55 (1983);  M.~J.~Duncan,
  Nucl.\ Phys.\  B {\bf 221}, 285 (1983);
   F.~Gabbiani and A.~Masiero,
  Nucl.\ Phys.\  B {\bf 322}, 235 (1989);
J.~S.~Hagelin, S.~Kelley and T.~Tanaka,
  Nucl.\ Phys.\  B {\bf 415} 293 (1994);
  F.~Gabbiani, E.~Gabrielli, A.~Masiero and L.~Silvestrini,
  Nucl.\ Phys.\  B {\bf 477}, 321 (1996).



\bibitem{ciuchini}
   M.~Ciuchini {\it et al.},
  JHEP {\bf 9810}, 008 (1998).



\bibitem{becirevic}
D.~Becirevic {\it et al.},
  Nucl.\ Phys.\  B {\bf 634}, 105 (2002).

\bibitem{utfit}
 M.~Bona {\it et al.}  [UTfit Collaboration],
  JHEP {\bf 0803}, 049 (2008).


\bibitem{dine}
 M.~Dine, R.~G.~Leigh and A.~Kagan,
  Phys.\ Rev.\  D {\bf 48}, 4269 (1993);
 R.~Barbieri, G.~R.~Dvali and L.~J.~Hall,
  Phys.\ Lett.\  B {\bf 377}, 76 (1996);
M.~C.~Chen and K.~T.~Mahanthappa,
  Phys.\ Rev.\  D {\bf 65}, 053010 (2002);
  S.~F.~King and G.~G.~Ross,
  Phys.\ Lett.\  B {\bf 574}, 239 (2003);
  G.~G.~Ross, L.~Velasco-Sevilla and O.~Vives,
  Nucl.\ Phys.\  B {\bf 692}, 50 (2004).

\bibitem{local}

K.~S.~Babu and S.~M.~Barr,
  Phys.\ Lett.\  B {\bf 387}, 87 (1996);
  K.~S.~Babu and R.~N.~Mohapatra,
  Phys.\ Rev.\ Lett.\  {\bf 83}, 2522 (1999).

  \bibitem{seiberg}

 P.~Pouliot and N.~Seiberg,
  Phys.\ Lett.\  B {\bf 318}, 169 (1993); D.~B.~Kaplan and M.~Schmaltz,
  Phys.\ Rev.\  D {\bf 49}, 3741 (1994);  L.~J.~Hall and H.~Murayama,
  Phys.\ Rev.\ Lett.\  {\bf 75}, 3985 (1995);  C.~D.~Carone, L.~J.~Hall and H.~Murayama,
  Phys.\ Rev.\  D {\bf 53}, 6282 (1996); P.~H.~Frampton and T.~W.~Kephart,
  Int.\ J.\ Mod.\ Phys.\  A {\bf 10}, 4689 (1995);
  T.~Kobayashi, S.~Raby and R.~J.~Zhang,
  Nucl.\ Phys.\  B {\bf 704}, 3 (2005);
  Y.~Kajiyama, E.~Itou and J.~Kubo,
  Nucl.\ Phys.\  B {\bf 743}, 74 (2006);
   M.~C.~Chen and K.~T.~Mahanthappa,
  Phys.\ Lett.\  B {\bf 652}, 34 (2007);
  I.~de Medeiros Varzielas, S.~F.~King and G.~G.~Ross,
  Phys.\ Lett.\  B {\bf 648}, 201 (2007).

 \bibitem{babu}
 K.~S.~Babu and J.~Kubo, Phys.\ Rev.\  D {\bf 71}, 056006 (2005).

\bibitem{kubolenz}
 N.~Kifune, J.~Kubo and A.~Lenz,
  Phys.\ Rev.\  D {\bf 77}, 076010 (2008).


\bibitem{kawamura}
Y.~Kawamura, H.~Murayama and M.~Yamaguchi, Phys.\ Rev.\  D {\bf 51}, 1337 (1995).



\bibitem{other}
S.~Pakvasa and H.~Sugawara,
  Phys.\ Lett.\  B {\bf 73}, 61 (1978);
  T.~Brown, N.~Deshpande, S.~Pakvasa and H.~Sugawara,
  Phys.\ Lett.\  B {\bf 141}, 95 (1984);
  E.~Ma,
  Phys.\ Rev.\  D {\bf 43}, 2761 (1991);
  P.~H.~Frampton and A.~Rasin,
  Phys.\ Lett.\  B {\bf 478}, 424 (2000);
  J.~Kubo, A.~Mondragon, M.~Mondragon and E.~Rodriguez-Jauregui,
  Prog.\ Theor.\ Phys.\  {\bf 109}, 795 (2003);
  E.~Ma, arXiv:hep-ph/0409075.



 \bibitem{othernu}
  E.~Ma and G.~Rajasekaran,
  Phys.\ Rev.\  D {\bf 64}, 113012 (2001);
  K.~S.~Babu, E.~Ma and J.~W.~F.~Valle,
  Phys.\ Lett.\  B {\bf 552}, 207 (2003);
   W.~Grimus, A.~S.~Joshipura, S.~Kaneko, L.~Lavoura and M.~Tanimoto,
  JHEP {\bf 0407}, 078 (2004); K.~S.~Babu and X.~G.~He,
  arXiv:hep-ph/0507217;
  G.~Altarelli and F.~Feruglio,
  Nucl.\ Phys.\  B {\bf 720}, 64 (2005);
  C.~Hagedorn, M.~Lindner and R.~N.~Mohapatra,
  JHEP {\bf 0606}, 042 (2006);
  E.~Ma, H.~Sawanaka and M.~Tanimoto,
  Phys.\ Lett.\  B {\bf 641}, 301 (2006);
  F.~Feruglio, C.~Hagedorn, Y.~Lin and L.~Merlo,
  Nucl.\ Phys.\  B {\bf 775}, 120 (2007); S.~F.~King and M.~Malinsky,
  Phys.\ Lett.\  B {\bf 645}, 351 (2007).


\bibitem{edm}
J.~R.~Ellis, S.~Ferrara and D.~V.~Nanopoulos,
  Phys.\ Lett.\  B {\bf 114}, 231 (1982); W.~Buchmuller and D.~Wyler,
  Phys.\ Lett.\  B {\bf 121}, 321 (1983); J.~Polchinski and M.~B.~Wise,
  Phys.\ Lett.\  B {\bf 125}, 393 (1983);  E.~Franco and M.~L.~Mangano,
  Phys.\ Lett.\  B {\bf 135}, 445 (1984); F.~del Aguila, M.~B.~Gavela, J.~A.~Grifols and A.~Mendez,
  Phys.\ Lett.\  B {\bf 126}, 71 (1983).


 \bibitem{nath}
 T.~Ibrahim and P.~Nath,
  Phys.\ Rev.\  D {\bf 57}, 478 (1998);
  M.~Brhlik, G.~J.~Good and G.~L.~Kane,
  Phys.\ Rev.\  D {\bf 59}, 115004 (1999);
 S.~Abel, S.~Khalil and O.~Lebedev,
  Nucl.\ Phys.\  B {\bf 606}, 151 (2001);
  J.~Hisano and Y.~Shimizu,
  Phys.\ Rev.\  D {\bf 70}, 093001 (2004); Phys.\ Lett.\  B {\bf 604}, 216 (2004);
For a review see M.~Pospelov and A.~Ritz,
  Annals Phys.\  {\bf 318}, 119 (2005).

 \bibitem{abel}
 S.~Abel, S.~Khalil and O.~Lebedev,
  Nucl.\ Phys.\  B {\bf 606}, 151 (2001).


\bibitem{weinberg}
S. Weinberg, in {\it Transactions of the New York Academy of Sciences} (New York Academy of Sciences, New York, 1977)  Ser. II, Vol.38, P.185; F. Wilczek and A. Zee, Phys. Rev. Lett. 42: 421, (1979);
H. Fritzsch, Phys. lett. B73: 317, (1978); Nucl. Phys. B155: 189, (1979).


\bibitem{hallrasin}
L.~J.~Hall and A.~Rasin,  Phys.\ Lett.\  B {\bf 315}, 164 (1993).


\bibitem{xing}
 Z.~z.~Xing, H.~Zhang and S.~Zhou, Phys.\ Rev.\  D {\bf 77}, 113016 (2008).


\bibitem{masip}
M.~Masip and A.~Rasin,
  Phys.\ Rev.\  D {\bf 52}, 3768 (1995); Nucl.\ Phys.\  B {\bf 460}, 449 (1996); Phys.\ Rev.\  D {\bf 58}, 035007 (1998).



\bibitem{carena}
  M.~S.~Carena, H.~E.~Haber, S.~Heinemeyer, W.~Hollik, C.~E.~M.~Wagner and G.~Weiglein,
  Nucl.\ Phys.\  B {\bf 580}, 29 (2000)

\bibitem{pdg}

 C.~Amsler {\it et al.}  [Particle Data Group],
 Phys.\ Lett.\  B {\bf 667}, 1 (2008).


\bibitem{he}
  N. G. Dashpande and X.-G. He, Phys. Rev. D49: 4812, (1994).



\bibitem{vempati}
M.~Ciuchini, A.~Masiero, P.~Paradisi, L.~Silvestrini, S.~K.~Vempati and O.~Vives,
Nucl.\ Phys.\  B {\bf 783}, 112 (2007).



\end{thebibliography}
\end{document}